\documentclass[12pt,preprint]{aastex}

\slugcomment{Accepted for publication in the Astrophysical Journal}
\shorttitle{}
\shortauthors{Maret et al.}

\begin{document}

\title{Spitzer mapping of molecular hydrogen pure rotational lines in
  NGC~1333: A detailed study of feedback in star formation}

\author{S\'ebastien Maret\altaffilmark{1}, Edwin A. Bergin}
\affil{Department of Astronomy, University of Michigan, 500 Church
  Street, Ann Arbor, MI 48109-1042}

\altaffiltext{1}{Present address: Laboratoire d'Astrophysique de
  Grenoble, Observatoire de Grenoble, Universit\'e Joseph Fourier,
  CNRS, UMR 571, BP~53, F-38041 Grenoble, France}

\author{David A. Neufeld}
\affil{Department of Physics and Astronomy, John Hopkins University,
  3400 North Charles Street, Baltimore, MD 21218}

\author{Joel D. Green, Dan M. Watson}
\affil{Department of Physics and Astronomy, University of Rochester,
  Rochester, NY 14627}

\author{Martin O. Harwit}
\affil{Department of Astronomy, Cornell University, Ithaca, NY
  14853-6801; and 511 H Street SW, Washington, DC 20024-2725}

\author{Lars E. Kristensen}
\affil{Leiden Observatory, P.O. Box 9513 NL-2300 RA Leiden, The
  Netherlands}

\author{Gary J. Melnick}
\affil{Harvard-Smithsonian Center for Astrophysics, 60 Garden
  Street, Cambridge, MA 02138}

\author{Paule Sonnentrucker} 
\affil{Department of Physics and Astronomy, John Hopkins University,
  3400 North Charles Street, Baltimore, MD 21218}

\author{Volker Tolls}
\affil{Harvard-Smithsonian Center for Astrophysics, 60 Garden Street,
  Cambridge, MA 02138}

\author{Michael W. Werner, Karen Willacy}
\affil{Jet Propulsion Laboratory, 4800 Oak Grove Drive, Pasadena, CA
  91109}

\and

\author{Yuan Yuan}
\affil{Department of Physics and Astronomy, John Hopkins University,
  3400 North Charles Street, Baltimore, MD 21218}

% \author{S\'ebastien Maret\altaffilmark{1,2}, Edwin
%   A. Bergin\altaffilmark{1}, David A. Neufeld\altaffilmark{3}, Joel
%   D. Green\altaffilmark{4}, Dan M. Watson\altaffilmark{4}, Martin O.
%   Harwit\altaffilmark{5}, Gary J. Melnick\altaffilmark{6}, Paule
%   Sonnentrucker\altaffilmark{3}, Volker Tolls\altaffilmark{6}, Michael
%   W. Werner\altaffilmark{7} and Karen Willacy\altaffilmark{7}}

% \altaffiltext{1}{Department of Astronomy, University of Michigan, 500
%   Church Street, Ann Arbor, MI 48109-1042}

% \altaffiltext{2}{Present address: Laboratoire d'Astrophysique de
%   Grenoble, Universit\'e Joseph Fourier, BP 53, F-38041 Grenoble,
%   France}

% \altaffiltext{3}{Department of Physics and Astronomy, John Hopkins
%   University, 3400 North Charles Street, Baltimore, MD 21218}

% \altaffiltext{4}{Department of Physics and Astronomy, University of
%   Rochester, Rochester, NY 14627}

% \altaffiltext{5}{Department of Astronomy, Cornell University, Ithaca,
%   NY 14853-6801; and 511 H Street SW, Washington, DC 20024-2725}

% \altaffiltext{6}{Harvard-Smithsonian Center for Astrophysics, 60 Garden
%   Street, Cambridge, MA 02138}

% \altaffiltext{7}{Jet Propulsion Laboratory, 4800 Oak Grove Drive,
%   Pasadena, CA 91109}

\begin{abstract}
  We present mid-infrared spectral maps of the NGC~1333 star forming
  region, obtained with the the \emph{Infrared Spectrometer} on board
  the \emph{Spitzer Space Telescope}. Eight pure H$_{2}$ rotational
  lines, from $S(0)$ to $S(7)$, are detected and mapped. The H$_{2}$
  emission appears to be associated with the warm gas shocked by the
  multiple outflows present in the region. A comparison between the
  observed intensities and the predictions of detailed shock models
  indicates that the emission arises in both slow (12 -- 24 km/s) and
  fast (36 -- 53 km/s) C-type shocks with an initial ortho-to-para
  ratio $\lesssim 1$.  The present H$_{2}$ ortho-to-para ratio
  exhibits a large degree of spatial variations. In the post-shocked
  gas, it is usually about 2, i.e. close to the equilibrium value
  ($\sim 3$). However, around at least two outflows, we observe a
  region with a much lower ($\sim 0.5$) ortho-to-para ratio. This
  region probably corresponds to gas which has been heated-up recently
  by the passage of a shock front, but whose ortho-to-para has not
  reached equilibrium yet. This, together with the low initial
  ortho-to-para ratio needed to reproduce the observed emission,
  provide strong evidence that H$_{2}$ is mostly in para form in cold
  molecular clouds.  The H$_{2}$ lines are found to contribute to 25
  -- 50\% of the total outflow luminosity, and thus can be used to
  ascertain the importance of star formation feedback on the natal
  cloud. From these lines, we determine the outflow mass loss rate
  and, indirectly, the stellar infall rate, the outflow momentum and
  the kinetic energy injected into the cloud over the embedded
  phase. The latter is found to exceed the binding energy of
  individual cores, suggesting that outflows could be the main
  mechanism for core disruption.
\end{abstract}


\section{Introduction}
\label{sec:introduction}

The early phases of the formation of a star are characterized by the
presence of bipolar outflows, which, as the material falls onto the
young star, eject matter at high speeds (several tens of km/s) to
distances up to tens of parsec away \citep[e.g.][]{Arce07}. These
outflows have a great impact on the surrounding gas. They can clear
out cavities in their natal molecular clouds, and may eventually cause
cloud disruption. As they propagate through molecular clouds, they
create shock waves that compress, heat up and alter the chemical
composition of the gas.  Theoretical studies of shock waves in the ISM
\citep[see][for a review]{Draine93} predict that their nature depends
on the shock velocity and the intensity of the magnetic field. Slow
shocks ($\le 45$ km/s for typical values of the magnetic field) are of
type C: the density, temperature and other quantities vary
continuously through their passage. Furthermore, they are
non-dissociative: molecules generally survive their passage, although
the chemical composition of the gas can be significantly altered. On
the other hand, faster shocks produce discontinuities in both the
density and temperature of the gas, and may dissociate molecules on
their passage.

Observations of H$_{2}$ pure rotational lines in shock regions can
provide important constraints on the physical and chemical conditions
that prevail in these regions. For example, the H$_{2}$ ortho-to-para
ratio (hereinafter \emph{opr}) is predicted to be greatly affected by
the passage of a C-type shock. In the cold (pre-shock) gas, H$_{2}$ is
expected to be mostly in para form \citep{Flower06a,Maret07a}. In the
shocked gas, reactive collisions with H atoms can convert p-H$_{2}$
into o-H$_{2}$ \citep{Timmermann98,Wilgenbus00}. This reaction has a
relatively high activation barrier ($\sim$ 4000 K), and therefore the
efficiency of the conversion depends critically on the temperature
that is reached in the shock. Indeed, \citet{Wilgenbus00} showed that
effective para to ortho H$_2$ conversion takes place in the
temperature interval $700-1300$~K, while \citet{Kristensen07} refined
this result in to the interval $800-3200$~K. In addition, the
efficiency of the conversion depends on the abundance of H atoms in
the gas. Fast shocks produce higher H abundances, which makes the
conversion faster.

Molecular hydrogen rotational lines have been observed in several star
forming regions with the \emph{Infrared Space Observatory} and, more
recently, with the \emph{Spitzer Space Telescope}
\citep{Werner04}. Using the \emph{Short Wavelength Spectrometer} on
board ISO, \cite{Neufeld98} observed five pure rotational lines (from
$0-0\, S(1)$ to $0-0\, S(5)$, hereafter simply referred as $S(1)$ and
$S(5)$, respectively) towards the Herbig-Haro 54 outflow
(HH~54). Using these multiple ortho and para lines, they determined
simultaneously the rotational temperature and the ortho-to-para
ratio. Interestingly, they measured an ortho-to-para ratio of 1.3,
which is significantly lower than the equilibrium value ($\sim 3$) at
the observed gas temperature ($\sim 650$ K). \cite{Lefloch03} used the
\emph{ISOCAM} camera to map the H$_{2}$ $S(2)$ to $S(7)$ towards the
HH~2 outflow, and found important spatial variations in the \emph{opr}
(between 1.2 and 2.5) around the HH~2 object. \cite{Neufeld06c} used
the \emph{Infrared Spectrometer} \citep[IRS,][]{Houck04} on board
\emph{Spitzer} to map the \emph{opr} towards HH~54 and HH~7-11, and
also observed a low, non-equilibrium \emph{opr}, with important
spatial variations. The fact that the \emph{opr} is out-of-equilibrium
suggests that the ortho-to-para conversion is relatively slow in these
regions: the observed \emph{opr} was interpreted as a fossil of an
earlier epoch when the gas was cooler.

This paper presents complete $5.2-36.5$~$\mu$m spectral maps obtained
with the IRS towards the NGC~1333 region in the Perseus cloud. The
observations cover a region of roughly $6\arcmin \times 10\arcmin$,
i.e. 0.4 pc $\times$ 0.5 pc assuming a distance of 220 pc
\citep{Cernis90}. To our knowledge, these are the largest complete
spectral maps obtained with the IRS. NGC~1333 contains many YSOs,
revealed by sub-millimeter \citep{Sandell01}, millimeter
\citep{Lefloch98} and mid-infrared continuum maps
\citep{Gutermuth08}. It also contains numerous molecular outflows,
detected for example in CO line emission \citep{Knee00}. Here we focus
on H$_{2}$ pure rotational line emission, from $S(0)$ to $S(7)$.
These lines are used to map the H$_{2}$ rotational temperature and
\emph{opr} over a large region, thus permitting detailed study of the
impact of outflows on the surrounding cloud.  The paper is organized
as follow. In \S \ref{sec:observations} we present the observations
and discuss the data reduction.  The spectra and maps that we obtained
are presented in \S \ref{sec:results}. In \S \ref{sec:analysis}, we
describe the analysis procedure that we used to derive the H$_{2}$
column density, rotational temperature and \emph{opr} from our
observations. In \S \ref{sec:discussion}, we compare our observations
with the predictions of shock models, we compare the variation of the
\emph{opr} with the temperature, and we discuss the impact of the
outflows on the cloud. Our conclusions are presented in \S
\ref{sec:conclusions}.

\section{Observations}
\label{sec:observations}

NGC~1333 was observed using the \emph{IRS} during Cycle 2 of the
General Observer program, in March and September 2006. The Long-High
(LH), Short-High (SH) and Short-Low (SL) modules of the IRS were used,
providing a complete spectral coverage from 5.2 to 36.5 $\mu$m. The
spectral resolution of these observations ($R = \lambda / \Delta
\lambda$) ranges from 64 to 128 for the SL mode, and is $\sim$600 for
the high resolution modules (SH and LH). The half-power beam size of
the instrument ranges from 3\arcsec\, at 5.2 $\mu$m to 10\arcsec\, at
38 $\mu$m \citep{Neufeld06c}.

The SH observations consist of nine different \emph{Astronomical
  Observations Requests} (AORs), each covering a rectangular region of
$\sim 142\arcsec \times 120\arcsec$. Each of these AORs was obtained
by moving the IRS slit in both perpendicular and parallel directions
of the slit length, with a half-slit width stepping in perpendicular
direction, and full-slit length in the parallel direction. Our SH
observations do not cover the tip of the HH~7-11 outflow; therefore we
have used archival data of this region obtained in February 2004 as
part of the IRAC Guaranteed Time program, and that consists of two
AORs of $58\arcsec \times 76\arcsec$ each. Details on these data can
be found in \citet{Neufeld06c}. When merged, the SH AORs cover a
region of $\sim 6\arcmin \times 10\arcmin$, roughly centered on SVS~13
($\alpha = 17^\mathrm{h} 22^\mathrm{m} 38.2^\mathrm{s}$ and $\delta =
-23 \degr 49 \arcmin 34.0 \arcsec$; J2000). The LH observations
consist of six AORs of $\sim 200\arcsec \times 273\arcsec$ each. As
for the SH observations, these were obtained with a half-slit width
stepping in perpendicular direction, and full-slit length in the
parallel direction. These observations cover a region of $\sim
8\arcmin \times 9\arcmin$. Finally, the SL observations consist of six
AORs of $\sim 202\arcsec \times 222\arcsec$, each obtained with a
full-slit width stepping in the perpendicular direction, and half-slit
width stepping in the parallel direction. These observations cover a
region of $\sim 6\arcmin \times 14\arcmin$.

All the data, including the archival ones, have been reduced using the
latest version (17.2) of the Spitzer Science Center pipeline. Further
reduction was performed using the SMART software package
\citep{Higdon04}, supplemented by the IDL routines described in
\cite{Neufeld06c}. These routines remove the bad pixels in the LH and
SH observations, extract spectra for each pixel along the slit of the
different modules, spatially re-sample the spectra on a regular grid
(with a spatial resolution corresponding to the pixel spacing,
i.e. 2.3\arcsec, 1.8\arcsec\, and 4.5\arcsec\, for the SH, SL and LH
modes, respectively), and finally produce continuum-subtracted
spectral maps by fitting a Gaussian for each spectral line. A first
look at our LH and SH spectral maps revealed the presence of stripes
in the direction perpendicular to the slit length. As discussed in
\cite{Neufeld07}, these stripes are due to the bad pixel extraction
routine, which interpolates missing pixel values in the dispersion
direction. Therefore, if a bad pixel is close to the central
wavelength of a given line, the intensity measured at a given position
along the slit is consistently underestimated, creating the observed
stripes. Following \cite{Neufeld07}, we have corrected this effect by
applying a correction factor for each line and each position along the
slit (see \citealt{Neufeld07} for details on how this correction
factor is determined). This technique was found to remove the stripes
from our maps quite efficiently. Note that our LH observations were
not affected by this problem because they were obtained with a
half-slit length stepping in the parallel direction, as opposed to the
SH and SL data which were obtained with a full-slit length stepping in
that direction. Therefore, in LH mode, any given position on the sky
was observed twice, and potential stripes were washed-out when
averaging the spectra together.
% Our final maps, in the FITS format,
% are made available on the web \emph{[ Details on the hosting of the
%   data to be figured-out ]}.

\section{Results}
\label{sec:results}

\subsection{Spectra}

\begin{figure}
  \centering \includegraphics[width=10cm]{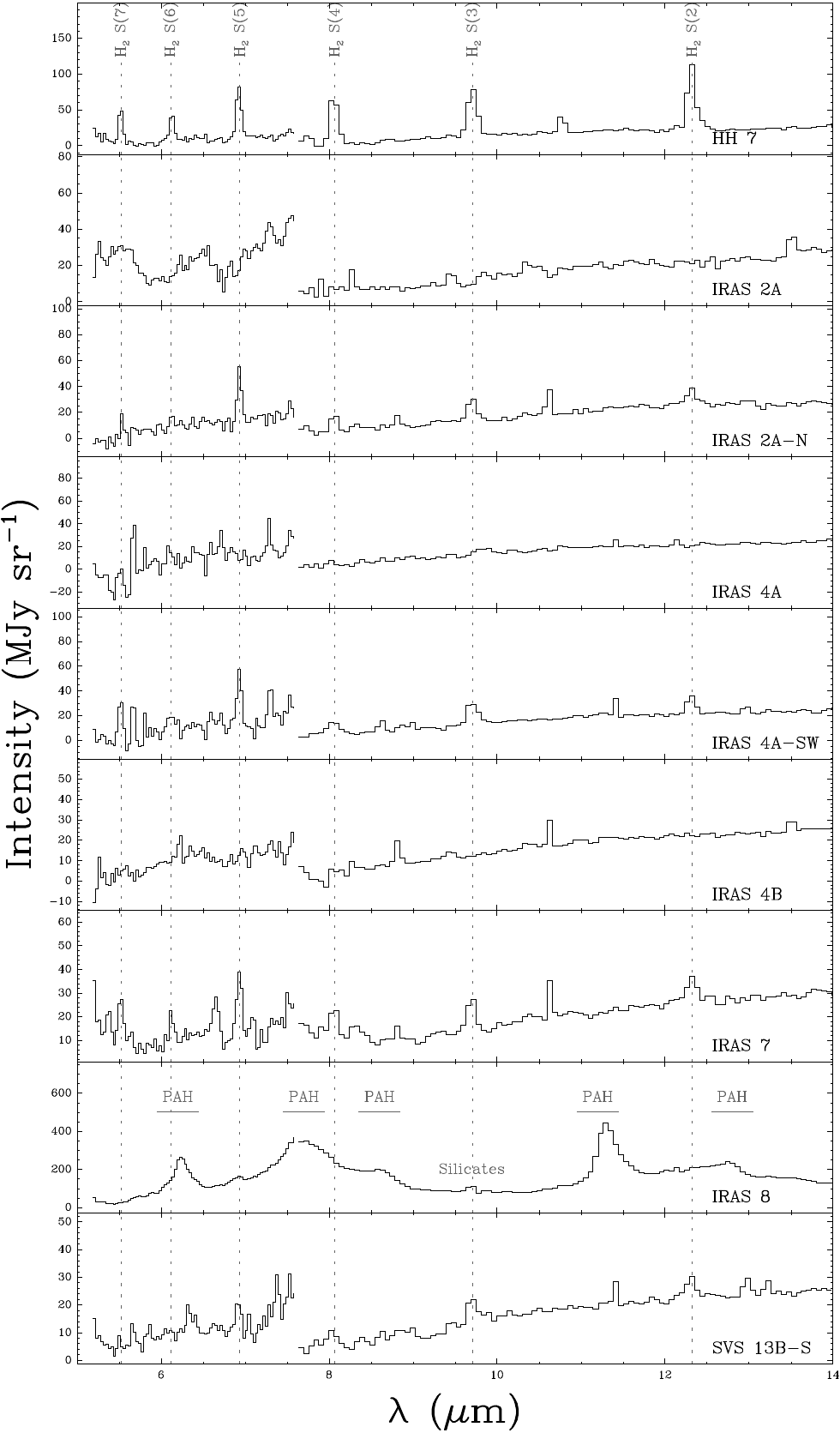}
  \caption{Averaged spectra observed with the SL module for 15\arcsec\
    Gaussian apertures centered, from top to bottom, on HH~7 ($\alpha
    = 03^\mathrm{h}29^\mathrm{m}08.45^\mathrm{s}$, $\delta =+31\degr
    15\arcmin29.2\arcsec$; J2000) IRAS~2A ($\alpha =
    03^\mathrm{h}28^\mathrm{m}55.59^\mathrm{s}$, $\delta
    =+31\degr14\arcmin37.3\arcsec$; J2000), IRAS~2A-S ($\alpha =
    03^\mathrm{h}28^\mathrm{m}54.04^\mathrm{s}$, $\delta
    =+31\degr13\arcmin31.2\arcsec$; J2000), IRAS~4A ($\alpha =
    03^\mathrm{h}29^\mathrm{m}10.29^\mathrm{s}$, $\delta
    =+31\degr13\arcmin31.8\arcsec$; J2000), IRAS~4A-SW ($\alpha =
    03^\mathrm{h}29^\mathrm{m}06.62^\mathrm{s}$, $\delta
    =+31\degr12\arcmin17.7\arcsec$; J2000), IRAS~4B ($\alpha =
    03^\mathrm{h}29^\mathrm{m}11.99^\mathrm{s}$, $\delta
    =+31\degr13\arcmin08.9\arcsec$; J2000), IRAS~7 ($\alpha =
    03^\mathrm{h}29^\mathrm{m}11.31^\mathrm{s}$, $\delta
    =+31\degr18\arcmin31.1\arcsec$; J2000), IRAS~8 ($\alpha =
    03^\mathrm{h}29^\mathrm{m}12.5^\mathrm{s}$, $\delta
    =+31\degr22\arcmin08\arcsec$; J2000), and SVS~13B-S ($\alpha =
    03^\mathrm{h}29^\mathrm{m}05.0^\mathrm{s}$, $\delta
    =+31\degr13\arcmin14\arcsec$; J2000). Pure rotational H$_{2}$
    transitions are indicated, as well as the 6.2, 7.2, 8.6, 11.2 and
    12.8 $\mu$m PAH features, and the 9.7 $\mu$m silicates absorption
    band\label{fig:sl-spectra}. The spectra between $\sim$ 5 and 7.5
    $\mu$m and between $\sim$ 7.5 and 14 $\mu$m correspond to the
    second and first order of the module, respectively.}
\end{figure}

\begin{figure}
  \centering \includegraphics[width=12cm]{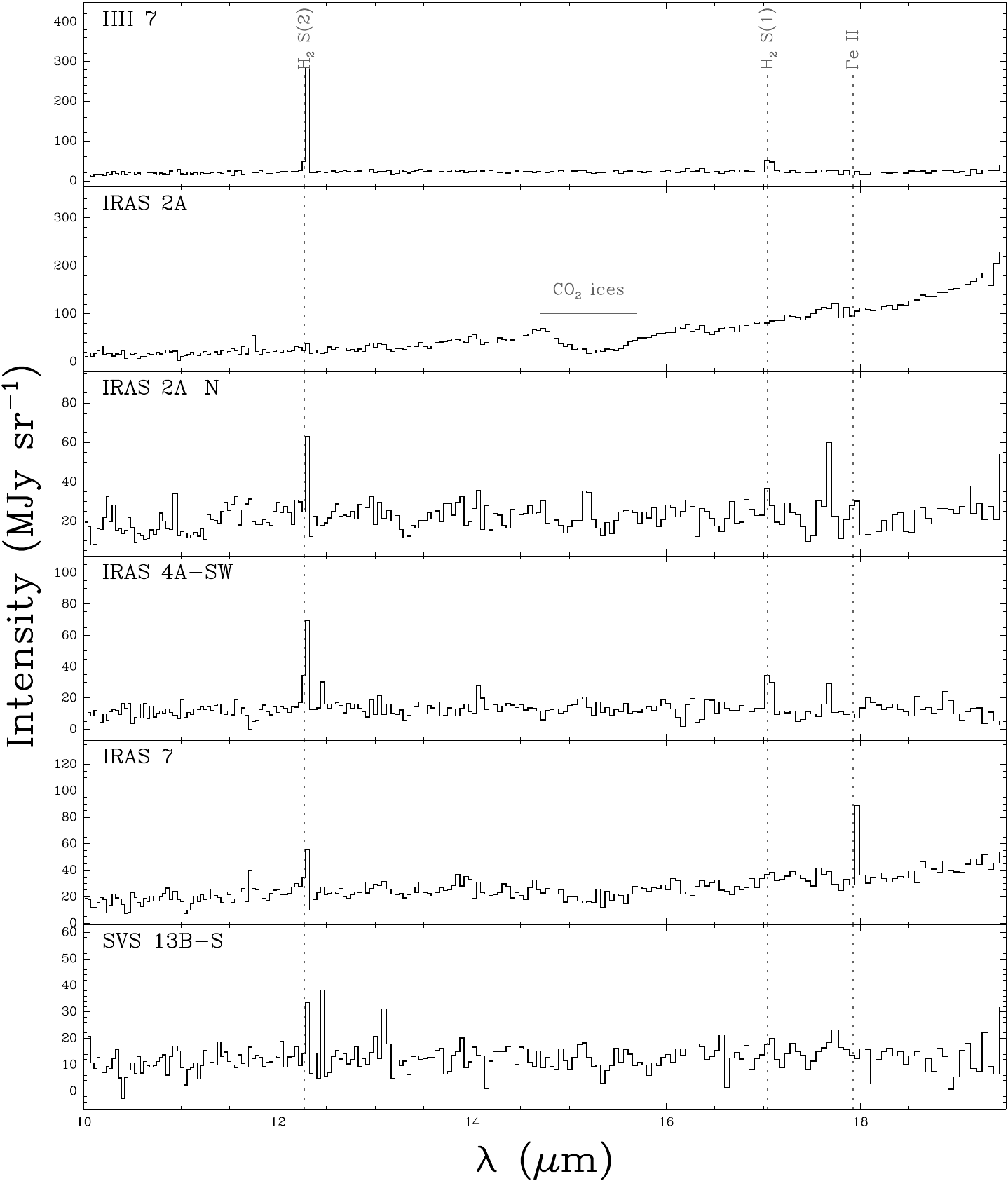}
  \caption{Same as Fig. \ref{fig:sl-spectra} for the SH module. The
    position of the 15.2 $\mu$ CO$_{2}$ ice feature is
    indicated\label{fig:sh-spectra}. For clarity the spectral
    resolution has been degraded by a factor 4.}
\end{figure}

\begin{figure}
  \centering \includegraphics[width=12cm]{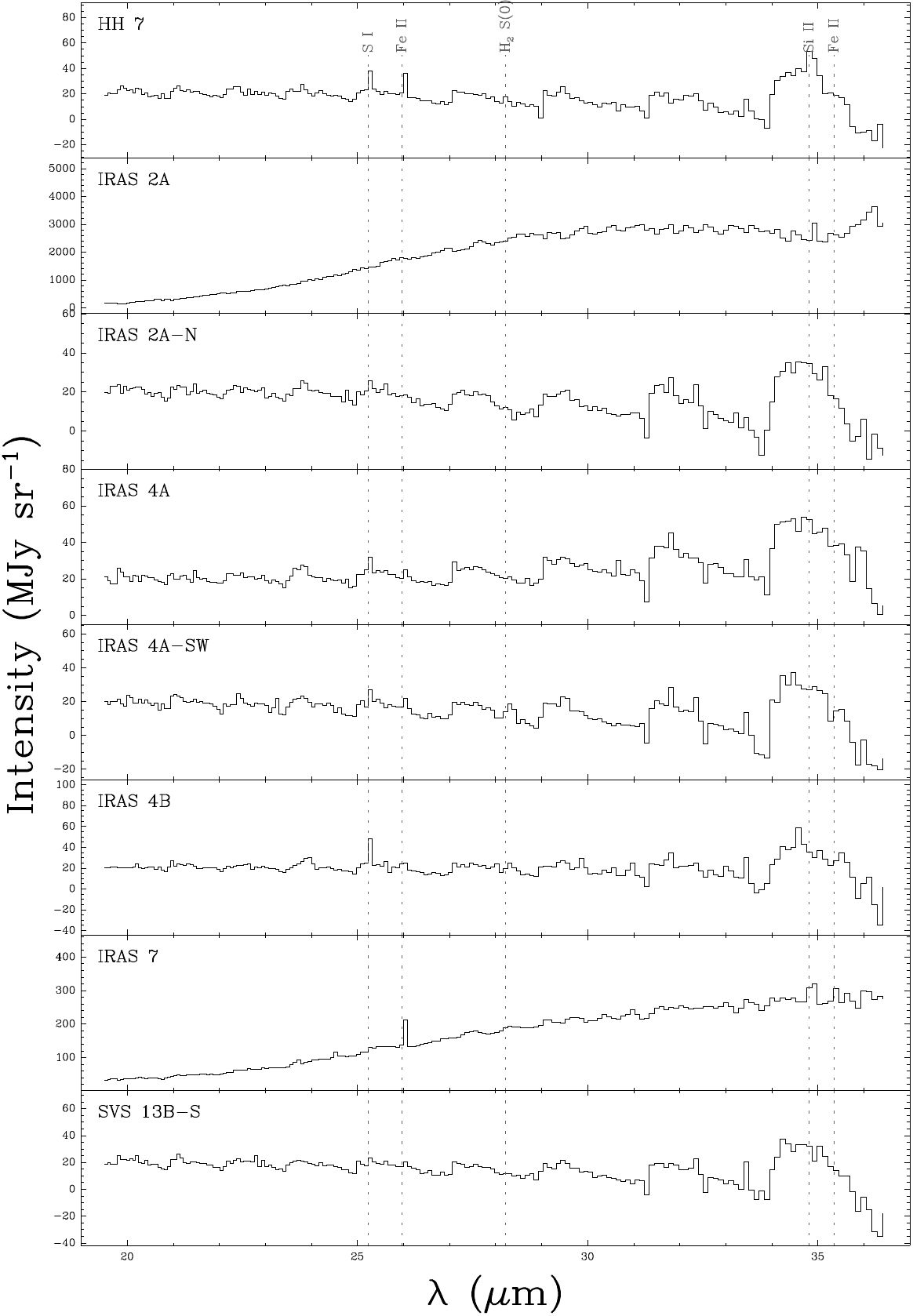}
  \caption{Same as Fig. \ref{fig:sl-spectra} for the LH module. For
    clarity the spectral resolution has been degraded by a factor
    4. The steps seen in some of these spectra are caused by offsets
    in response of the different orders of the
    module.\label{fig:lh-spectra}}
\end{figure}

Fig. \ref{fig:sl-spectra} to \ref{fig:lh-spectra} present average
spectra obtained on several positions of our maps. These spectra were
obtained by averaging all the spectra observed within a 15\arcsec\,
FWHM Gaussian aperture around a given position, in order to increase
the signal-to-noise ratio. We detect eight pure rotational lines --
from $S(0)$ to $S(7)$ -- towards HH~7. On the other hand, no H$_{2}$
emission is detected towards the NGC~1333-IRAS~4A and 4B Class 0
protostars. Several fine structure atomic lines are also detected,
such as the \ion{Fe}{2} $^{4}F_{7/2} - ^{4}F_{9/2}$ (17.9 $\mu$m), the
\ion{S}{1} $^{3}P_{1} - ^{3}P_{2}$ (25.2 $\mu$m), \ion{Fe}{2}
$^{6}D_{7/2} - ^{6}D_{9/2}$ (26.0 $\mu$m) and \ion{Si}{2}
$^{2}P_{3/2}^{0} - ^{2}P_{1/2}^{0}$ (34.8 $\mu$m) lines. Table
\ref{tab:intensities} gives the intensities measured towards several
positions for these lines. Thanks to an overlap in the wavelength
range covered by the SL and SH modules, the H$_{2}$ $S(2)$ intensity
was measured with each module, and was found to agree within $\sim
20\%$.

Several PAH features -- at 6.2, 7.2, 8.6, 11.2 and 12.8 $\mu$m -- as
well as the 9.7 $\mu$m silicates absorption band are clearly detected
towards IRAS~8. Finally, the CO$_{2}$ ice feature at 15.2~$\mu$m is
seen on IRAS~2. A discussion of all the detected spectral features is
beyond the scope of this paper; in the following, we focus on the
interpretation of the H$_{2}$ line emission, and we postpone the
discussion of other lines and spectral features to forthcoming papers.

\subsection{Maps}

Fig. 4~a-i show maps of the H$_{2}$ $S(0)$ to $S(7)$ lines. H$_{2}$
$S(1)$, $S(2)$, $S(3)$, $S(4)$ and $S(5)$ emission is readily detected
along several outflows. The H$_{2}$ $S(0)$, $S(6)$ and $S(7)$ maps
have a lower signal-to-noise ratio, and emission of these lines is
barely detected at this spatial resolution (4.5 and 1.8\arcsec\, for
the LH and SL modes, respectively). In the following, we discuss the
morphology of the H$_{2}$ emission along the different outflows.

\subsubsection{SVS~13 and HH~7-11}
\label{sec:svs-13-hh7-11}

The Class I source SVS~13 \citep{Strom76} is associated with a chain
of Herbig-Haro objects, HH~7-11. These objects are excited by a
high-velocity molecular outflow, which is detected in CO and SiO
rotational lines \citep{Bachiller98b, Bachiller00b, Knee00}, as well
as the H$_{2}$ $1-0$ $S(1)$ ro-vibrational line at 2.12 $\mu$m
\citep{Aspin94,Hodapp95,Khanzadyan03}. A second outflow, roughly
orientated along a north-south axis and driven by the SVS~13B Class 0
source, is detected in SiO $J = 2-1$ emission \citep{Bachiller98c}.

\begin{figure}
  \figurenum{4~a}
  \centering \includegraphics[]{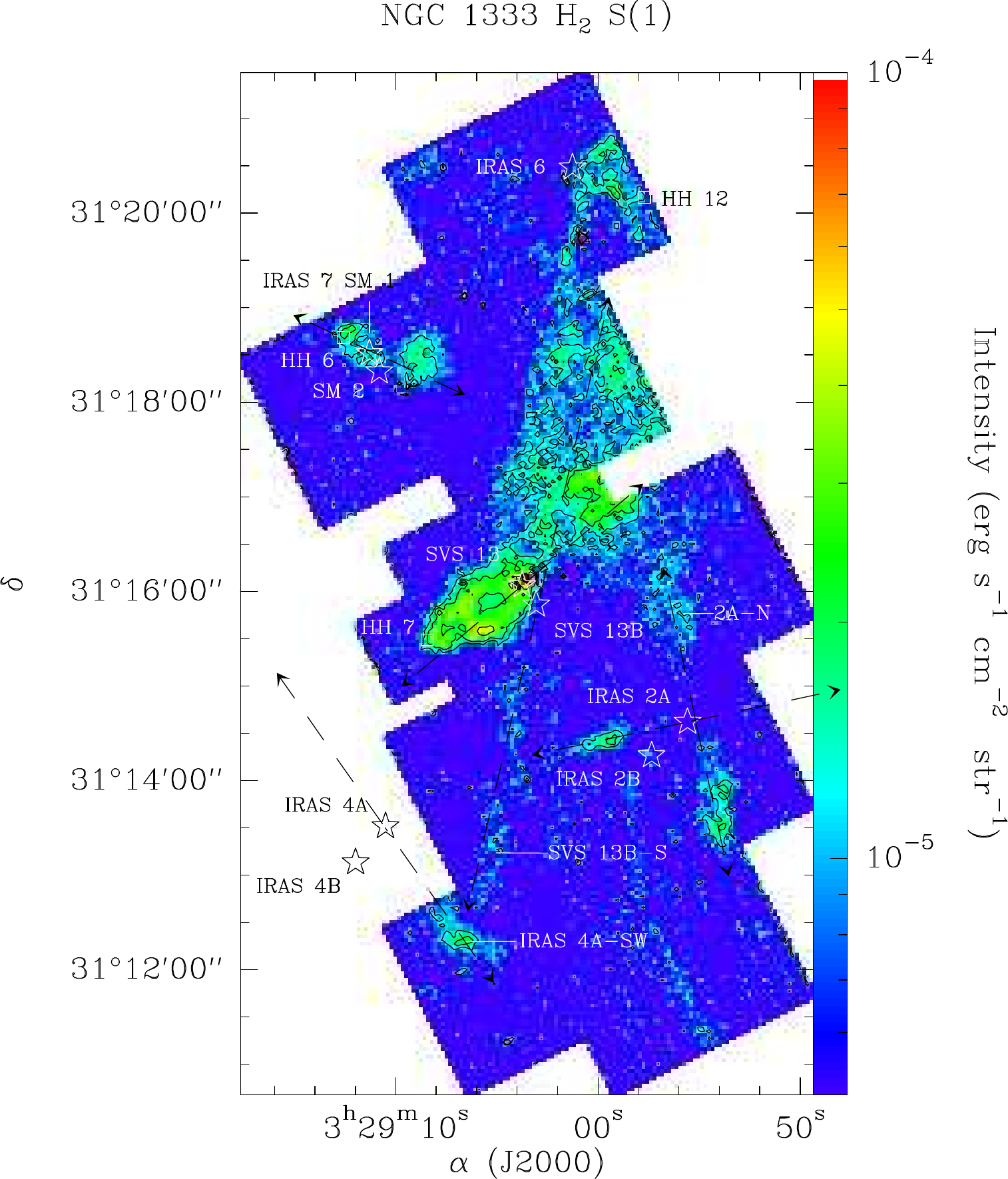}
  \caption{H$_{2}$ $S(1)$ continuum subtracted map obtained with the SH module. YSOs are
    indicated by open stars, while several HH objects are indicated by
    open squares. The outflows discussed in the text are indicated by
    dashed arrows. The two positions along the IRAS~2 and IRAS~4A
    outflows that are discussed in the text, IRAS~2A-N and IRAS~4A-SW,
    are also shown. Contours show the 3, 5 and 10
    $\sigma$ noise levels. \emph{See the electronic edition of the Journal for
      panels d-i.}
    \label{fig:h2-s1-map}}
\end{figure}

\begin{figure}
  \figurenum{4~b}
  \centering \includegraphics[]{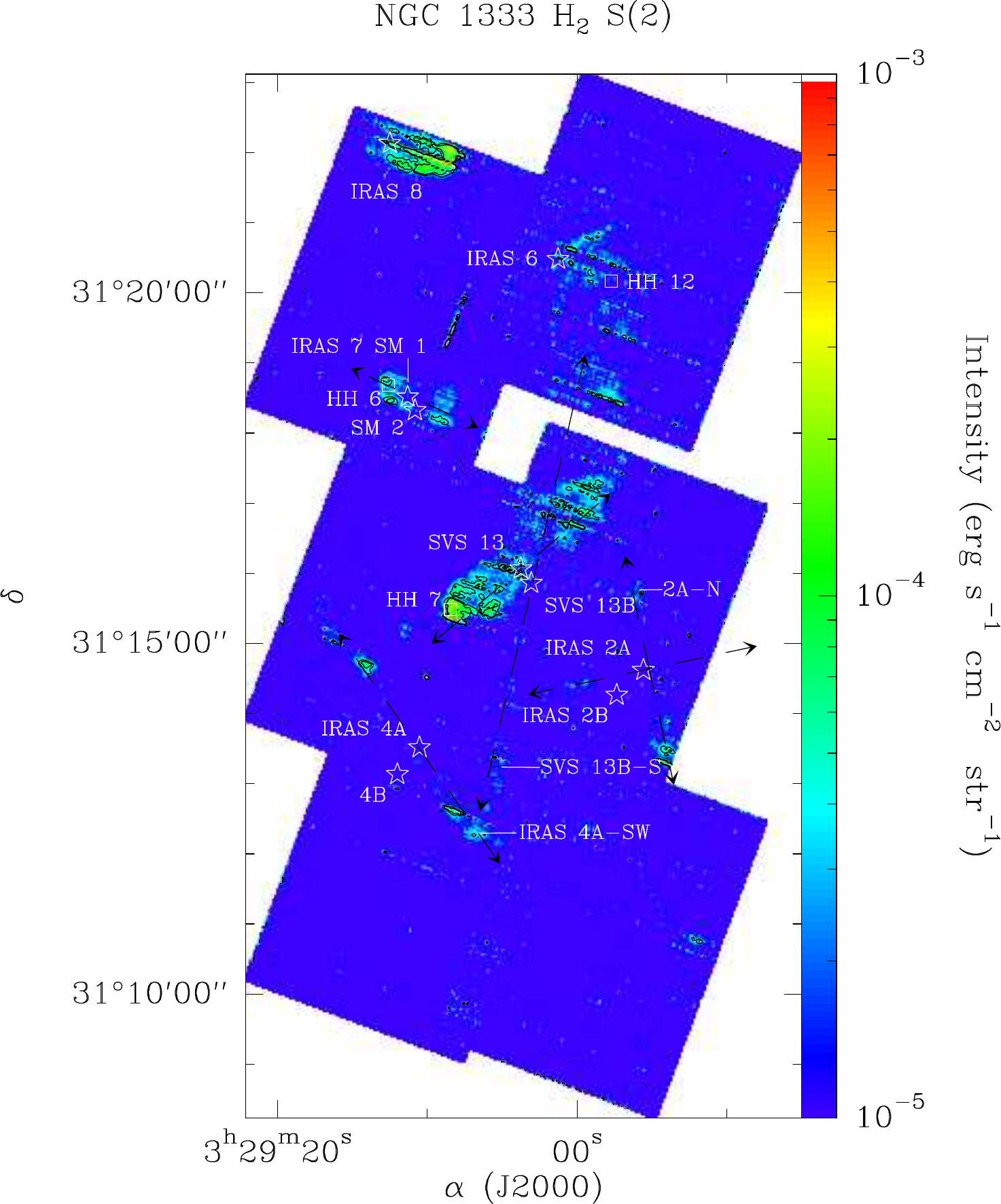}
  \caption{H$_{2}$ $S(2)$ map obtained with the SL module. Contours
    show the 3 and 5 $\sigma$ noise level.\label{fig:h2-s2-map}}
\end{figure}

\begin{figure}
  \figurenum{4~c}
  \centering \includegraphics[]{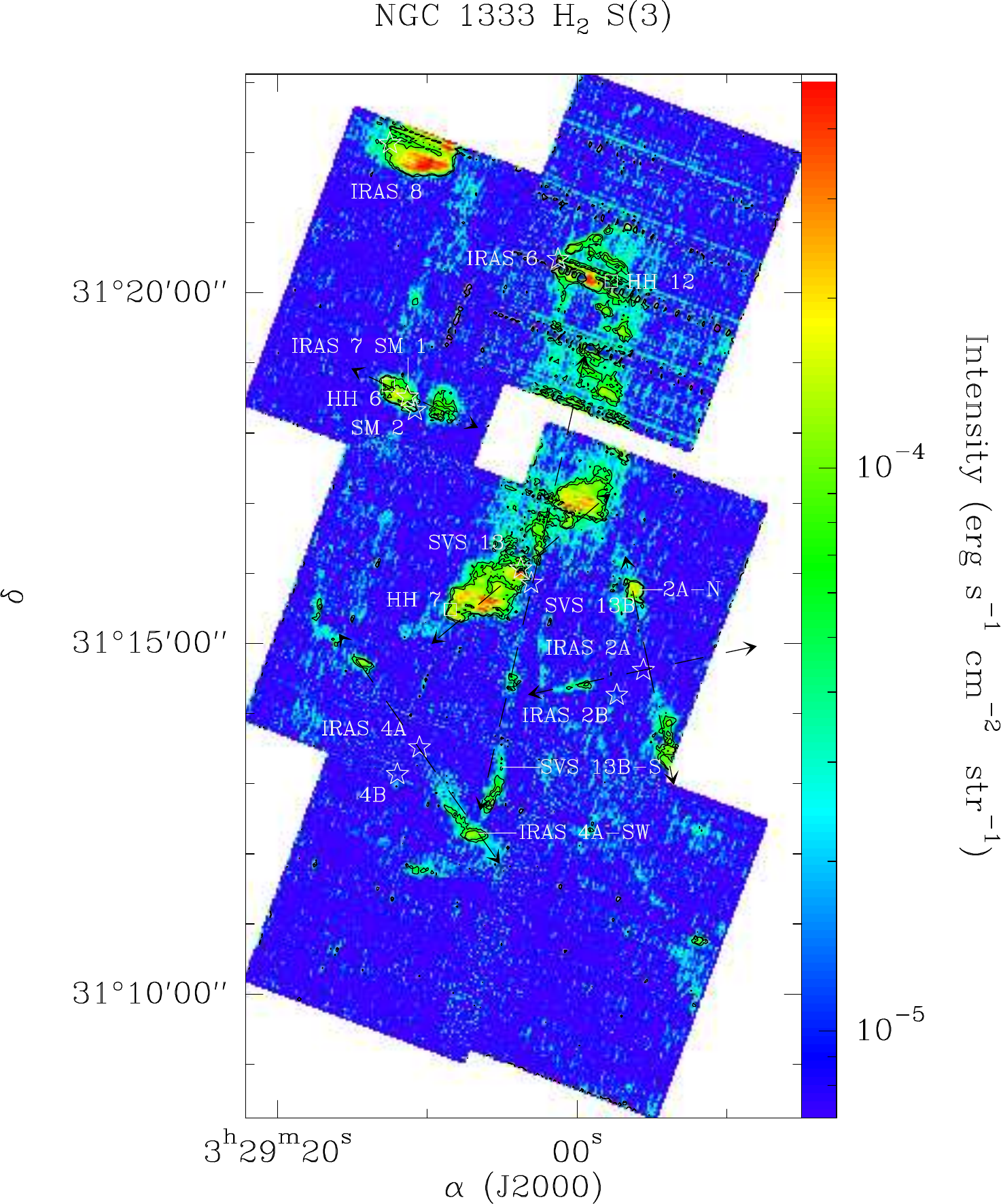}
  \caption{H$_{2}$ $S(3)$ map obtained with the SL module. Contours
    show the 3 and 5 $\sigma$ noise level.\label{fig:h2-s3-map}}
\end{figure}

The south-east lobe of the HH~7-11 outflow is clearly detected in our
$S(1)$ map (Fig. \ref{fig:h2-s1-map}). This map does not cover the
north-west lobe, but emission from this lobe is seen in the $S(2)$ and
$S(3)$ maps (Fig. \ref{fig:h2-s2-map} and \ref{fig:h2-s3-map}).
H$_{2}$ $S(1)$ and $S(3)$ emission is also detected along the
north-south outflow originating from SVS~13B. The position angle (from
the north to east) of the H$_{2}$ emission is about $160\degr$.
% is slightly different from the one
% of SiO emission ($180\degr$; \citealt{Bachiller00b}), which may
% indicate a precession of the outflow axis. 
The emission extends as far as $\sim 3\arcmin$ south of SVS~13B, up to
the tip of the NGC~1333-IRAS~4A outflow (see below). Some weak
emission along this flow is also seen in the H$_{2}$ $S(2)$
line. Diffuse $S(1)$ emission extends to the north along the same
axis, and may be associated with this outflow as well (see \S
\ref{sec:ngc1333-iras6-hh12}).

\subsubsection{NGC~1333-IRAS~2}

NGC~1333-IRAS~2 \citep[][hereafter IRAS~2]{Jennings86} is a binary
system composed of the IRAS~2A and IRAS~2B Class 0 protostars. This
system is associated with two outflows that both originate
$\sim6\arcsec$ west of IRAS~2A \citep{Knee00}. The first one is
roughly orientated along a north-south axis, and is detected in CO
\citep{Liseau88, Sandell94a} and H$_{2}$ $1-0$ $S(1)$
\citep{Hodapp95}. The second outflow is roughly oriented along an
east-west axis (position angle $\sim 104\degr$) and is detected in CO,
CH$_{3}$OH and SiO
\citep{Sandell94a,Bachiller98b,Knee00,Jorgensen04b}.

In our maps, we observe a peak of $S(1)$, $S(2)$ and $S(3)$ emission
east of IRAS~2B along the axis of the east-west outflow. The position
of this peak corresponds to the peak of the CS and CH$_{3}$OH
emission, and presumably arises from a bow-shock
\citep{Langer96,Bachiller98b}.  No emission is seen west of the
source, but our maps have limited coverage in that direction. Two
H$_{2}$ emission spots are also seen north and south of IRAS~2A. These
two peaks lie at a position angle of $\sim 12\degr$, consistent with
the orientation of the H$_{2}$ $1-0$ $S(1)$ jet, but slightly
different from the position angle of the CO jet \citep[$\sim
25\degr$][]{Sandell94a}. No emission is seen towards the protostars.

\subsubsection{NGC~1333-IRAS~4}

NGC~1333-IRAS~4 is also a binary system, composed of two Class 0
protostars, IRAS~4A and IRAS~4B. IRAS~4A drives a powerful and well
collimated outflow that extends roughly over $4\arcmin$, and is
detected in H$_{2}$ $1-0$ $S(1)$ \citep{Hodapp95} as well as CO and
SiO lines \citep{Blake96,Knee00}. A more compact outflow, with an
inclination close to 90$\degr$ (i.e. almost perpendicular to the plane
of the sky), is driven by IRAS~4B \citep{Blake96}.

We detect H$_{2}$ $S(1)$ emission at the southwestern tip of the
IRAS~4A outflow. Unfortunately, our SH maps do not cover other parts
of the flow. However, $S(2)$ and $S(3)$ emission peaks are seen at to
the north-east and south-west of IRAS~4A. No H$_{2}$ is detected from
the IRAS~4B outflow, nor from IRAS~4A and IRAS~4B protostars
themselves.

\subsubsection{NGC~1333-IRAS~7 and HH~6}

NGC~1333-IRAS~7 is associated with two sub-millimeter sources, SM~1
and SM~2 \citep{Sandell01}. The former is coincident with the cm
source VLA~27 \citep{Rodriguez99}, as well as with a water maser
\citep{Henkel86}. A molecular outflow, oriented along a east-west axis
and probably driven by SM~1, is detected in CO
\citep{Liseau88,Knee00}. This outflow is the driving source for the
chain of HH objects located east of SM~1, which include HH~6.  Our
maps reveal bipolar H$_{2}$ emission orientated along the same axis as
the CO outflow.

\subsubsection{NGC~1333-IRAS~6 and HH~12}
\label{sec:ngc1333-iras6-hh12}

NGC~1333-IRAS~6 is coincident with a PMS star (SVS~107) and the
continuum cm source VLA~42 \citep{Rodriguez99}. An elongated structure
is also seen towards that source in the sub-mm maps of
\citet{Sandell01}. west of NGC~1333-IRAS~6 is a chain of HH objects
that include HH~12. \cite{Knee00} suggested that these objects are
excited by the outflow originating from SVS~13B.

In our H$_{2}$ maps we see some diffuse emission west of IRAS~6 that
extends from the south down to the north west lobe of the SVS~13
outflow. Although this diffuse emission appears broadly aligned with
the axis of the SVS~13B, it is difficult to say from the present data
if the H$_{2}$ emission arises in the shocked gas from this outflow.

\subsubsection{NGC~1333-IRAS~8}

NGC~1333-IRAS~8 is part of the optical nebulosity north of the cloud
(see Fig. \ref{fig:h2-s2-map}). It is associated with a PMS star,
SVS~3. Although the IRS spectrum is dominated by PAH features (see
Fig. \ref{fig:sl-spectra}), we also detect extended H$_{2}$ $S(2)$,
$S(3)$ and $S(5)$ (but no $S(4)$) to the south east of the source.

\section{Analysis}
\label{sec:analysis}

\subsection{Rotational diagrams}
\label{sec:rotational-diagrams}

In Fig. \ref{fig:rotational-diagram}, we show rotational diagrams
obtained for 15\arcsec\ FWHM Gaussian apertures centered at several
positions along the outflows: HH~7, IRAS~4A-SW, IRAS~2A-S, IRAS~7 and
SVS~13B-S. These positions are indicated on the H$_{2}$ maps
(Fig. 4~a-i).  Line intensities were corrected for extinction using
the standard ($R_{v} = 3.1$) galactic dust opacity curve from
\citet{Weingartner01}, assuming $A_{v}$ = 2 for all sources (the value
measured towards HH~8; \citealt{Gredel96}). In each diagram and for
each line, we have plotted $ln (N_{u} / (g_u g_s))$, where $N_{u}$ is
the column density in the upper level, $g_u$ is the rotational
degeneracy (equal to $2 \,J + 1$) and $g_s$ is the spin degeneracy (3
for ortho, and 1 for para transitions). If the ortho-to-para ratio is
equal to its high temperature limit (\emph{i.e.} 3), and if H$_{2}$
transitions are thermalized at a single temperature, then the observed
values would line up on a straight line. Instead, the diagram show a
``zig-zag'' pattern, indicating that the ortho-to-para ratio is lower
than 3. In addition, the rotational diagrams show a positive
curvature; the observed values cannot be fit with a single excitation
temperature. This suggests that the observed H$_{2}$ emission arises
in a mixture of gas with different temperatures, as already observed
in HH~7 and HH~54 by \citet{Neufeld06c}. Following these authors, we
have fitted the observations with two gas components (herein after the
``warm'' and the ``hot'' components). Note that this two temperature
model is an approximation; the gas has more likely a continuous
temperature distribution. The total H$_{2}$ column density,
ortho-to-para ratio and excitation temperature in each gas component
were kept as free parameters, and were adjusted to fit the
observations using a minimization routine.  These three parameter can
be constrained quite independently from each other. The ortho-to-para
ratio is constrained from the ``zig-zag'' pattern in the rotational
diagram. The rotational temperature is determined from the slope of
the rotational diagram. Finally, the total H$_{2}$ column density is
determined from the ordinate of the points in the diagram.  This
procedure produced good fits to the data; in particular, no clear
departures from the LTE is seen, even for the highest energy lines.
The critical densities -- above which collisional excitation dominates
over radiative de-excitation, and energy levels are populated
according to Boltzmann's statistic -- range between 5 and $4 \times
10^{5}$ cm$^{-3}$ for the $S(0)$ and $S(7)$ lines respectively
\citep[assuming a kinetic temperature of 1000
K;][]{LeBourlot99}. Therefore the highest energy lines ($S(6)$ and
$S(7)$) must arise in relatively dense gas ($> 10^{5}$ cm$^{-3}$).

The best fit parameters for HH~7, IRAS~4A-SW, IRAS~2A-S, IRAS~7,
SVS~13B-S are given in Table \ref{tab:rotational-diagram-results}. The
rotational temperature towards these sources is between $\sim$ 300 and
$\sim$ 600 K for the warm component, and $\sim$ 1000 - 1500 K for the
hot component. We measure an \emph{opr} of 0.3 -- 0.7 and 1.4 -- 2.1
in the warm and hot components, respectively\footnote{The best fit
  parameters we obtain for HH~7 differ slightly from those reported by
  \citet{Neufeld06c} on the same source. The reason is that we used
  new data in addition to those of \citet{Neufeld06c}.}. For both
components, the \emph{opr} is much lower than its equilibrium value
for the measured (rotational) gas temperatures; for kinetic
temperatures higher than 300 K, the equilibrium value is essentially
3. For all sources, we measure a higher \emph{opr} in the hot
component than in the warm component. Similar behavior was noted
previously by \cite{Neufeld06c,Neufeld07}. The dependence of
\emph{opr} on temperature is discussed in further detail below. The
H$_{2}$ column densities are typically $\sim 2 - 6 \times 10^{19}$
cm$^{2}$ and $\sim 3 - 6 \times 10^{18}$ cm$^{2}$ in the warm and hot
components, respectively. Thus, the column density of the hot
component is more than an order of magnitude smaller than the column
density of the warm component.

\subsection{H$_{2}$ column density, rotational temperature and
  ortho-to-para ratio maps}
\label{sec:h_2-column-density}

Thanks to their large spatial coverage, our observations essentially
map the ortho-to-para ratio, rotational temperature, and column
density of H$_{2}$ in the core of the NGC~1333 cloud. For this
purpose, the H$_{2}$ line maps were re-sampled on a common grid with a
5$\arcsec$ spacing. Only regions of the sky that were observed with
all three modules were considered. We then constructed a rotational
diagram in each pixel of the resulting map. As mentioned above, the
H$_{2}$ rotational emission arises from at least two gas
components. With three free parameters for each component (column
density, \emph{opr} and rotational temperature), we need at least six
lines detected in a given pixel of the maps. In practice, few pixels
in our maps meet this criterion. Therefore, when constructing the
maps, only the H$_{2}$ $S(0)$ to $S(3)$ transitions were considered,
which allows for greater spatial coverage of the region.  Since the
$S(0)$ to $S(3)$ line emission is dominated by the warm component (see
Fig. \ref{fig:rotational-diagram}), the column density, \emph{opr} and
rotational temperature derived using these lines correspond to that
component.

On Fig. \ref{fig:h2-column-density-map},
\ref{fig:h2-rotational-temperature-map} and \ref{fig:h2-opr-map} we
provide maps of the H$_{2}$ column density, rotational temperature and
\emph{opr}. These quantities could be measured along both lobes of the
SVS~13 outflow and around IRAS~7. We could also map those at the
south-west tip of the IRAS~4A outflow, as well at the northern,
southern and western tips of the IRAS~2 outflows. Elsewhere in the
map, we did not detect the H$_{2}$ $S(0)$, $S(1)$, $S(2)$ and $S(3)$
lines simultaneously, or the lines were detected with an insufficient
signal-to-noise ratio. Where measured, some important variations in
these quantities are seen. The H$_{2}$ column density ranges from $3
\times 10^{18}$ to $10^{20}$ cm$^{-2}$. The highest column densities
are seen along the HH~7-11 outflow (see
Fig. \ref{fig:h2-column-density-map}). The rotational temperature
ranges between 200 and 650 K, with a maximum reached close to the
position of HH~7, as well as the IRAS~2N position (see Fig.
\ref{fig:h2-rotational-temperature-map}). Finally, the H$_{2}$
ortho-to-para ratio ranges from 0.1 and 3 (Fig. \ref{fig:h2-opr-map}).

% Figures \ref{fig:h2-rotational-temperature-opr-map-svs13-iras4} and
% \ref{fig:h2-rotational-temperature-opr-map-iras2-iras7} show a
% side-by-side comparison between the rotational temperature and the
% \emph{opr} around HH~7-11, IRAS~4, IRAS~2 and IRAS~7. 
Some important variations in both the rotational temperature and the
\emph{opr} are seen. Along the HH~7-11 outflow, between SVS~13 and
HH~8, we measure a rotational temperature of $\sim$ 400 K and an
\emph{opr} ranging between 1.5 and 2. Around HH~7, we measure a higher
rotational temperature ($\sim$ 650 K), but a lower \emph{opr} ($\sim$
0.5). The presence of hot gas with a low \emph{opr} in front of the
HH~7-11 has already been observed by \cite{Neufeld06c}. They suggested
that this gas has been compressed and heated-up by the passage of the
shock, but the ortho-to-para ratio has not reached LTE yet. Along the
northwestern lobe of this flow, the spectra are noisier, and large
variations from one pixel to the next. Around IRAS~4-SW, the
rotational temperature and the \emph{opr} were measured only in a few
pixels making the comparison between these two quantities
difficult. On the other hand, we observe some variation in the
rotational temperature along the north and south lobes of the IRAS~2
outflow. Like in HH~7-11, the rotational temperature is higher at the
tip of this outflow. However, no corresponding variation is seen in
the \emph{opr}. Some relatively hot gas ($\sim$ 500 K) with a low
\emph{opr} ($\sim 1$) is also observed around the eastern lobe of the
IRAS~7 outflow. The variation of the \emph{opr} as a function of the
\emph{$T_{rot}$} are further discussed in
Appendix~\ref{sec:variation-opr-with}.
 
\section{Model and Discussion}
\label{sec:discussion}

\subsection{Comparison of the observed emission with the predictions
  of a shock model}
\label{sec:comp-observ-emiss}

In this section, we compare the observed H$_{2}$ line emission towards
several outflow positions with the predictions of a stationary, planar
C-type shock model that includes ortho para interconversion
\citep{Flower03}. The shock model has several free parameters: the
pre-shock density $n$, the pre-shock velocity $v$, the initial
(pre-shock) ortho-to-para ration $opr_{i}$, and the intensity of the
transverse magnetic field. The latter is expressed as $b = B_{t} \,
n_{0}^{-1/2}$, where $B_{t}$ is the transverse magnetic field in
$\mu$G and $n_{0}$ is the pre-shock density in cm$^{-3}$. The model
has been run for a grid of parameters (Kristensen et al. in prep), and
the results of these computations are available on the
web\footnote{\url{http://www.strw.leidenuniv.nl/$\sim$kristensen/models}}.

In Fig. \ref{fig:shock-models}, we compare the column densities
obtained from our observations on aperture averaged positions with
those predicted by the shock model. As a first approach, we have tried
to reproduce the observed column densities with a single component
shock. The value of $b$ was set to 1, and the other parameters of the
model were adjusted until a good agreement with the observations was
obtained. We found that the observed $S(0)$, $S(1)$, $S(2)$ emission
were fairly well fitted for a pre-shock density of $10^{4}$ cm$^{-3}$,
an initial \emph{opr} of 0.01 or 1, a pre-shock velocity of 12--24
km/s, and a beam filling factor $ff$ of typically 10\%. However, this
model greatly underestimates the observed lines with higher
energies. In other words, the temperature reached in a slow shock is
not sufficient to explain the $S(3)$ to $S(7)$ line that we
observe. This emission requires higher gas temperatures that are
reached only in faster shocks. Therefore we have tried to match the
observed column densities with a combination of slow and fast shock
components. Both the pre-shock density and initial \emph{opr} were
kept the same as for the slow shock component, and the velocity of the
fast shock component was adjusted in order to match the $S(3)$ to
$S(7)$ lines. We found that the emission of these lines were well
matched for a shock velocity of 36--53 km/s and a filling factor of
typically 1\%. In fact, this two component model provides an excellent
fit to the observations (see Fig. \ref{fig:rotational-diagram}), with
the exception of the $S(0)$ line [$E_\mathrm{up} = 510$~K] in IRAS-2,
IRAS-7 and SVS13B-S, which is somewhat underestimated by the
model. Possibly, this line arises in a separate, non-shock component
towards these sources. Indeed, in their study of supernovae remnants,
\citet{Neufeld07} found that the $S(0)$ emission was more extended
than other spectral lines they observed. In one source, they found
evidence for both a shock excited $S(0)$ component and a diffuse
emission component.

Table \ref{tab:shock-model-parameters} gives the best-fit model
parameters for the slow and fast shock components. Towards all
positions, we obtain a good fit for a pre-shock density of $10^{4}$
cm$^{-3}$. The shock velocity in the slow shock component is between
12 and 24 km/s, while that in the fast shock component is between 36
and 53 km/s. For all sources but IRAS~4-SW, the emission is well
matched assuming an initial \emph{opr} of 0.01. For the latter, we
obtain a slightly better fit for an initial \emph{opr} of
1.0. However, the grid of models covers only a relatively small sample
of $opr_i$ (0.01, 1, 2 and 3) so this parameter is not constrained
precisely. It is likely that $opr_i$ values ranging between 0.01 and 1
would also provide a good fit to the observations. The filling factor
ranges from 2 to 15\% for the slow component, and from 0.8 to 3\% for
the fast component. Since shocks driven by a collimated jet are not
planar but have a bow-shape, it seems natural that we observe a range
of velocities: at the apex of the bow, the shock is moving faster with
respect to the wings of the bow \citep{Smith91}. This high-velocity
component has a small beam filling factor, because the high velocities
are present only at the apex of the bow. On the other hand, slower gas
-- leading to lower excitation -- is present in the broader wings of
the bow.

It is worth noting that the fit is not be unique. For example, we
found that the emission observed in HH~7 can also be fitted by a model
with an higher pre-shock density (10$^{5}$ cm$^{-3}$), but a smaller
pre-shock velocity (15 and 30 km/s for the slow and fast component,
respectively). In this case, the observations require an initial
\emph{opr} of 1, because the para to ortho conversion at these shock
velocities is not efficient enough. Another reasonable fit was found
for a pre-shock density of 10$^{6}$ cm$^{-3}$ and pre-shock velocities
of 10 and 20 km/s for the slow and fast component,
respectively. Therefore the pre-shock density and the pre-shock
velocity are difficult to constrain simultaneously from the present
observations.  However, other constraints on these parameters have
been obtained using FIR lines observations.  \citet{Molinari00}
observed CO and H$_{2}$ lines with ISO, and compared these
observations with the predictions of \cite{Kaufman96} shock
models. They found that the observed emission was well reproduced by a
shock with a velocity ranging between 15 and 20 km/s, and a pre-shock
density of 10$^{4}$ cm$^{-3}$. \citet{Giannini01} also compared the
H$_{2}$O and CO line emission observed with ISO-LWS towards the
red-lobe of the IRAS~4 outflow to the model predictions of
\cite{Kaufman96}. Their observations are consistent with a velocity
ranging between 15 and 20 km/s, and a pre-shock density of 10$^{4}$
cm$^{-3}$ (see their Fig. 8). Thus the pre-shock density adopted here
is consistent with the values derived by these authors using CO
emission, which is a much more sensitive probe of density. It is also
interesting to note that the shock velocity range they obtain agree
well with what we obtain here for the slow shock
component. Furthermore \citet{Smith03} obtained images of the
$2-1~S(1)$ and $1-0~S(1)$ H$_{2}$ emission lines around the HH~7 bow
shock. They found that their observations were well reproduced by a
C-type shock model with a pre-shock density of $8 \times
10^{3}$~cm$^{-3}$, a bow speed of 55~km/s and a magnetic field of
97~$\mu$G (corresponding to $b \simeq 1$ for that pre-shock
density). These parameters agree remarkably well with the values
obtained here. The bow speed is slightly higher than the velocity we
derive for the fast shock component (45~km/s). This is consistent with
models that predict that pure rotational H$_{2}$ lines arise in the
wings of the bow, where the shock velocity is smaller than at the apex
\citep[see][Fig~5]{Smith03}.

An important result of the present study is that the H$_{2}$
observations cannot be reproduced by an initial ortho-to-para ratio
greater than 1; models with $opr_i = 2$ or 3 produce ``flat''
rotational diagrams, i.e. \emph{opr} values that are close to the high
temperature limit of 3. This indicates that H$_{2}$ in the cold gas
(i.e. before the passage of a shock) is mostly in para form.
% Faster shocks tends to overproduce the $S(2)$ and $S(3)$ emission with
% respect to the $S(0)$ and $S(1)$. In the rotational diagram this
% translates as a slope which is shallower than observed. This is simply
% because faster shocks heats up the gas to higher temperature than
% slower ones. Conversely, slower shocks produces curves in the
% rotational diagram with slopes that stepper than the observations. In
% addition, faster shocks produce higher \emph{opr}, because they
% heat-up the gas to higher temperatures and produce more atomic H
% (which reacts with p-H$_{2}$ to form o-H$_{2}$, as mentioned earlier
% in the paper). Thus faster shocks produce curves with more ``zig-zag''
% in the rotational diagram then slower ones. Of course, the amount of
% ``zig-zag'' is also controlled by the initial \emph{opr}.
\citet{Flower06a} studied the evolution of the \emph{opr} as a
function of time in a pre-stellar core. In their computations, H$_{2}$
is assumed to form on the grains with an \emph{opr} of 3. Reactive
collisions between o-H$_{2}$ and p-H$_{3}^{+}$ (or o-H$_{3}^{+}$)
forms p-H$_{2}$, and consequently the \emph{opr} decreases with
time. The steady state value ($\sim 3 \times 10^{-3}$ for $T = 10$ K,
and $\sim 3 \times 10^{-2}$ for $T = 30$ K) is reached after $t \sim 3
\times 10^{7}$ yr, for a density of $10^{4}$ cm$^{-3}$. Our
observations and modelling suggest that the initial ortho-to-para
ratio is lower than 1. According to the \citet{Flower06a}
computations, an ortho-to-para of 1 is reached after $3 \times 10^{6}$
year (see their Fig. 1); this is comparable to the life-time of nearby
molecular clouds obtained by \citet{Hartmann01}. Therefore the value
of the initial \emph{opr} is consistent with that is expected from
chemical models.

% The disagreement between the observations and the predictions of a
% single C-type stationary shock model may suggest some departure from
% the stationary state. \citet{Chieze98} studied the temporal evolution
% of MHD shocks, and found that for C-type shocks, the stationary state
% is reached after $10^{5}$ yr. This is much longer than the dynamical
% times of outflows, which are typically of a few hundred
% years. Therefore, C-type shocks are unlikely to have reached the
% stationary state.  \citet{Chieze98} predict that before reaching the
% stationary state, the shock retains both C and J-type characteristics;
% first, a discontinuity appear in the neutral gas component, similarly
% to what happens in a J-type shock. A magnetic precursor (with no
% discontinuity) is present in the pre-shock gas, as in a C-type
% shock. Finally, the ion-neutral coupling causes the discontinuity in
% the ionized component to progressively disappear, and the shock
% finally become of C-type. \citet{Flower99} computed the H$_{2}$
% rotational emission in a non-stationary shock, and found that at early
% time, the J-type component dominate the emission, while at later time,
% the C-type component dominate the emission. At intermediate times
% (typically 10$^{3}$ yr), both components are found to contribute to
% this emission, similarly to what we observe here. A comparison between
% our observations with such time dependent shock model predictions
% could be useful to constrain the age the outflows.

\subsection{Thermal history of the shocked gas}

The measure of the \emph{opr} along the several outflows in the cloud
provides constraints on the thermal history of the gas. Our
observations and models suggest that in the cold gas, H$_{2}$ is
mostly in para form. Shock waves heat up the gas and trigger
para-to-ortho conversion by reactive collisions with H, so the
\emph{opr} increases. However, the reaction has a large activation
barrier ($\sim$3900~K), so it is only efficient for gas temperatures
greater than a few thousand Kelvins. As the post-shock gas cools down,
the interconversion becomes less efficient, and the \emph{opr} remains
frozen. Of course, reactions with H$_{3}^{+}$ will tend to decrease
the \emph{opr} in the post-shock gas. However, this occurs on
relatively long timescales \citep[$\sim 10^{6}$ yr for $n_\mathrm{H} =
10^{4}$;][]{Flower06a}, when compared to the dynamical timescale of
the flows.  Thus the \emph{opr} retains a memory of the thermal
history of the gas.

Constraints on the thermal history of the gas may be placed by simple
considerations. If we assume that a parcel of the gas is warmed-up by
a passage of a shock wave to a temperature $T$ for a duration $\tau$,
and that ortho-para interconversion occurs only by reactive collisions
with H, then the fraction of H$_{2}$ in ortho form in this parcel is
given by \citep{Neufeld06c}:

% \begin{equation}
%   \frac{opr (\tau)}{1 + opr (t)} = \frac{opr_{i}}{1 + opr_{i}} e^{-n(\mathrm{H})\, k\, \tau} + 
%   \frac{opr_{\mathrm{LTE}}}{1 + opr_{\mathrm{LTE}}} (1 - e^{-n(\mathrm{H})\, k\, \tau})
%   \label{eq:1}
% \end{equation}

\begin{equation}
  f_\mathrm{ortho} (\tau) = f_{\mathrm{ortho}}^i \, \mathrm{exp}
  \left( \frac{-n(\mathrm{H})\, k\, \tau}
    {f_\mathrm{ortho}^\mathrm{LTE}} \right) +
  f_\mathrm{ortho}^\mathrm{LTE} \left[ 1 - \mathrm{exp} \left(
      \frac{-n(\mathrm{H})\, k\, \tau} {f_\mathrm{ortho}^\mathrm{LTE}}
    \right) \right]
  \label{eq:1}
\end{equation}

\noindent
where $f_{\mathrm{ortho}}^i$ is the initial ortho fraction,
$f_\mathrm{ortho}^\mathrm{LTE}$ is the ortho fraction at the local
thermodynamic equilibrium, $n(\mathrm{H})$ is the hydrogen density,
and $k = k_{po} + k_{op}$ is the sum of the rate coefficients for
para-to-ortho and ortho-to-para conversion, given by $k = 10^{-11}\,
\mathrm{exp}\,(-3900/T)$ cm$^{3}$~s$^{-1}$
\citep{Schofield67}\footnote{In the analogous equation given by
  \citep{Neufeld06c}, $k_{po}$, the rate coefficient for para-to-ortho
  conversion alone, was erroneously given in place of $k$.  The former
  is 25\% smaller than the latter (a difference which is negligible
  relative to other uncertainties).}. Following \cite{Neufeld06c}, we
assume that H in the shocked gas is mainly produced by the formation
of water from atomic oxygen, so that $n(\mathrm{H})/n(\mathrm{H}_2)
\sim 10^{-3}$. Thus, for a pre-shock H$_{2}$ density of 10$^{4}$
cm$^{-3}$, $n(\mathrm{H}) \sim 10$~cm$^{-3}$. Fig. \ref{fig:opr-time}
shows the \emph{opr} as a function of time for different gas
temperatures, assuming an initial \emph{opr} of 0.01. In HH~7, we
measure an \emph{opr} of 0.37 and 1.99 in the warm ($T = 611$~K) and
hot ($T=1401$~K) components, respectively. This corresponds to
timescales of $\sim 8 \times 10^{3}$ and $\sim \times 10^{3}$~yr for
the warm and hot components, respectively.

The values we obtain can be compared to the prediction of shock
models. The timescale during which the gas temperature is elevated on
passing through a non dissociative shock is given by $N_{s} / (n_{0}
\, v_{s})$, where $N_{s}$ is the column density of the warm shocked
gas, $n_{0}$ is the pre-shock density, and $v_{s}$ is the pre-shock
velocity \citep{Neufeld06c}. Assuming a pre-shock density of
10$^{4}$~cm$^{-3}$, and a pre-shock velocity of 20~km~s$^{-1}$, we
obtain, using the analytical expression for $N_{s}$ from
\cite{Neufeld06c}, a shock timescale of 400~yr.
% As already mentioned, the predicted width of the region of the shock
% where the temperature is greater than 1000~K is $\sim 10^{3}$ AU, for
% a shock velocity of $20$ km~s$^{-1}$. Dividing the shock width by the
% shock velocity, we obtain a time during which the gas is hotter than
% 1000~K of $\sim 250$ yr.
This is about an order of magniture lower than the timescales we
obtain for both the warm and hot components of HH~7. This may suggest
that several shocks waves have passed through the quiescent gas and
have progressively increased the \emph{opr}, so $opr_i$ is greater
than 0.01. If we assume that $opr_i$ is 0.37 -- the value we observe
in the warm component of HH~7, at the tip of the HH~7-11 outflow -- we
found, using Eq. (\ref{eq:1}), that 800~yr are needed to reach the
value of 1.99 measured in the hot component, assuming a temperature of
1401~K. This is broadly consistent with the 400~yr timescale obtain
above.

% This is still about an order of magnitude larger than the
% predicted time during which the gas temperature is hotter than 1000~K;
% thus, a different value for $opr_i$ alone can not explain the
% differences in the timescales.

% This simple analysis indicates that efficient para-to-ortho conversion
% must occur in parts of the shock that are hotter than a few thousands
% Kelvins; at lower temperatures, the shock wave propagates too fast for
% the conversion to occur. Indeed, our shock model predicts maximum
% temperatures of $\sim$1500~K and $\sim$5500~K for the slow and fast
% shock components in HH~7. If we assume that $opr_i$ = 0.01, we found
% that $\sim$500~yr and $\sim$200~yr are needed to reach the \emph{opr}
% values observed in the warm (0.37) and hot (1.99) component,
% respectively. This is in reasonable agreement with the 250~yr value
% obtained above.

\subsection{Global outflow properties}
\label{sec:glob-outfl-prop}
 
We have obtained 5.2 -- 35.8 $\mu$m spectroscopic maps encompassing
nearly the full spatial extent of 5 molecular outflows from embedded
young protostars (HH~7-11, SVS~13B, IRAS~4A, IRAS~2, IRAS~7).  In the
following section we will explore how the information obtained in
these maps provides a direct measurement of the mass loss rate and
outflow energetics from these young stars.
 
\subsubsection{Determination of total outflow energy loss}
 
As shown in Fig. 4~a-i, we have detected optically thin emission from
eight rotational transitions of H$_2$ within these flows.  Using these
data, we can calculate the total H$_2$ luminosity in rotational
emission, $L_{\rm H_2}$, in the outflow by summing the emission over
all rotational states and at each position within the region
associated with each outflow.  The transitions we have observed
account for most of the luminosity carried by H$_2$ rotational
emissions, the brightest transitions falling within the range with
$J_{up} = 2 - 9$ covered by \emph{Spitzer-IRS}.  We provide the value
of L$_{\rm H_2}$ for each flow in
Table~\ref{tab:outflow-properties}. The H$_2$ luminosity is related to
the total outflow cooling rate by $L_{\rm tot} = \frac{1}{f_c} L_{\rm
  H_2}$, where $f_c$ is the fraction of the total cooling due to H$_2$
rotational emissions.

The shock models of \cite{Kaufman96} can be used to explore the
contribution of H$_2$ to the overall cooling of the outflow via
emission.  Based on analysis of the H$_2$ emission
(\S~\ref{sec:comp-observ-emiss}) we find typical shock parameters of
$n_0 \sim 10^4$ cm$^{-3}$ and $v_s \sim 10-50$ km/s.  Under these
conditions \citet{Kaufman96} find that $f_c \sim 0.2 - 0.7$.  The
majority of H$_2$ cooling is via rotational lines, with only few
percent contribution from H$_2$ vibrational emission.  The rest of the
cooling flux arises from rotational, and to a lesser extent
vibrational, lines of H$_2$O and CO, along with [\ion{O}{1}] fine
structure emission.
    
The \emph{Infrared Space Observatory} has observed the primary cooling
lines of CO, [\ion{O}{1}], H$_2$O, OH, and H$_2$ with the SWS and LWS
spectrometers in some of our sources. \citet{Molinari00} observed HH~7
and compile the luminosities of the major cooling lines within the
bandpass.  For the HH~7 shock they find that $4.8 \times
10^{-2}$~L$_{\odot}$ is released by \ion{O}{1}, $2 \times
10^{-2}$~L$_{\odot}$ by CO, $0.7 \times 10^{-2}$~L$_{\odot}$ by
H$_2$O, and $2.3 \times 10^{-2}$~L$_{\odot}$ by H$_2$.  However, the
\ion{O}{1}, CO, and H$_2$O observations were obtained using the LWS
instrument with an 80$''$ beam centered on HH~7, while the H$_2$
observations covered only the $S(1)-S(5)$ transitions within a 14$''
\times 20''$ SWS beam.  Our observations in Fig.~4a-i demonstrate that
the H$_2$ emission extends well beyond the SWS aperture and the ISO
SWS value is therefore a lower limit.  We have convolved our data with
a simulated 80$''$ LWS beam and have recomputed the total luminosity
of H$_2$ from $S(0)-S(7)$.  We find that the H$_2$ luminosity in an
80$''$ beam centered on HH~7 is 0.1~L$_{\odot}$. Note that in this
source, the luminosity of the H$_{2}$~$1-0~S(0)$ ro-vibration line is
$3 \times 10^{-4}$~L$_{\odot}$ \citep{Khanzadyan03}, well below the
H$_2$ luminosity in rotational emission. Combining the ISO
observations with this revised value, we find that the H$_2$ provides
about 50\% of the total cooling ($f_c \sim 0.5$), in good agreement
with models.

Information for other flows can be gleaned from the
summary of ISO cooling lines of \cite{Giannini01} which included two
of our sources: the IRAS~2 north-south flow, and the IRAS~4A outflow.
For these sources, we find (after adapting the results of Giannini et
al. to our adopted distance of 220 pc) that $f_c \sim 0.25$.  Using
this information the total cooling luminosity of the outflows in
NGC~1333 is approximately $L_{tot} \sim 2-4\; L_{\rm H_2}$.  For
HH~7-11 we will adopt a $f_c = 0.5$ and for all other flows (which are
not as prominent as HH~7-11) we adopt $f_c = 0.25$.

\subsubsection{Determination of the stellar mass loss by outflows}
 
The total energy loss generated by the outflow driven shock determined
above is a direct measurement of the mechanical luminosity of the
outflows.  In this fashion, $\frac{1}{2}\dot{M}_o v_s^2 = (1-
f_m)L_{tot} = (1 - f_m)\frac{1}{f_c} L_{\rm H_2}$, where $(1 - f_m)$
is the fraction of shock mechanical energy that is translated into
internal excitation, as opposed to translational motion.
\cite{Kaufman96} estimate that $(1 - f_m) \sim 0.75$ and we can then
use the H$_2$ cooling emission to determine the outflow mass loss
rate, $\dot{M}_o$ from the young star.  This equation can be
simplified and placed in terms of the observed H$_2$ luminosity and
estimated shock velocity:

\begin{equation}
  \dot{M}_o \simeq 10^{-6} (1- f_m) \frac{1}{f_c} \left(\frac{L_{\rm H_2}}{10^{-2}
      \;L_{\odot}}\right) \left(\frac{v_s}{10\;{\rm km/s}}\right)^{-2}\;
  {\rm M_{\odot}\; yr}^{-1}.
\end{equation}

\noindent
To determine the shock velocity, we use the models of H$_2$ emission
discussed in \S~\ref{sec:comp-observ-emiss}.  In general, the data is
best fit by a combination of fast and slow shocks with the slow shock
encompassing a larger filling factor within the beam.  We have
therefore estimated the average shock velocity for each flow by using
a filling factor weighted average of the shock velocity based upon the
model fits.  In Table~\ref{tab:outflow-properties} we provide the
observed H$_2$ luminosity, assumed shock velocity, and derived outflow
properties.

In general we find that the outflows from the Class 0 protostars in
the NGC~1333 cloud have values $\dot{M}_w$ $\sim 0.6 - 2\times
10^{-6}$ M$_\odot$ yr$^{-1}$.  These values are comparable to values
estimated in the literature
\cite[e.g.][]{Bontemps96,Giannini01,Hatchell07}.  In some cases
measurements include sources within our sample.  For example,
\citet{Giannini01} use FIR cooling lines towards IRAS~4 to derive
$\dot{M}_w = 0.4 - 1.4 \times 10^{-6}$ M$_\odot$ yr$^{-1}$ where we
estimate $\dot{M}_w = 2 \times 10^{-6}$ M$_\odot$ yr$^{-1}$.  The
derived mass outflow rates can be related to the mass accretion rate
($\dot{M}_a$) using theoretical models of outflow generation.  While
the details and results can vary, a typical number is $\dot{M}_o
\simeq 0.1 \; \dot{M}_a$ \citep{Pudritz07}, and our estimates provide
$\dot{M}_a \sim 10^{-5}$ M$_\odot$ yr$^{-1}$.  In this regard,
\citet{DiFrancesco01} independently derived a mass infall rate of $1
\times 10^{-4}$ M$_\odot$ yr$^{-1}$ towards IRAS~4A using P Cygni
spectral line profiles of H$_2$CO.  In addition, \cite{Maret02}
derived an accretion rate of $5 \times 10^{-5}$ M$_\odot$ yr$^{-1}$ by
modelling the FIR water lines emission from IRAS~4A and 4B
envelopes. This is in reasonable agreement with our estimate of $2
\times 10^{-5}$ M$_\odot$ yr$^{-1}$ for this source.

It is useful to place our results in a broader context.  Previous
estimates of these quantities for molecular outflows have generally
used velocity-resolved CO emission to provide a measure of the total
H$_2$ mass in the flow, and used the velocity extent of the lines to
trace the kinematics \citep[e.g.][]{Lada85,Bontemps96}.  However,
there are a number of uncertainties inherent in this method.  The CO
abundance has been calibrated to H$_2$ in some clouds
\citep[e.g.][]{Frerking82}, but can vary if some CO remains frozen on
grains \citep[e.g.][]{Bergin07b} or is photodissociated by radiation
generated in the shock or leaking through the outflow cavity.  In
addition, the CO molecules can be excited in both the low temperature
natal core and in the warmer outflow.  Thus towards the central core
the spectral line contains a mixture of emission from the high
velocity dispersion core and low dispersion core.  To some extent
these components can be disentangled, but there is significant
uncertainty regarding the outflow mass present at low velocities that
also correspond to the core \citep{Lada96}.  CO emission itself is
often optically thick, even in the extended line wings.  Observations
of isotopologues are therefore required, but not always available to
correct for the effects of optical depth \citep[e.g.][]{Yu99,Arce01b}.
Finally, estimates of the CO velocities are dependent on the outflow
inclination.  In a handful of cases, this can be directly measured if
proper motion data can be used to measure the tangential velocity to
compare to the observed radial velocity in the spectral lines.  In the
majority of cases this information is not available, and inclinations
have instead been estimated by models of the flow velocity at
different inclinations compared to the observed emission distribution
\citep[e.g.][]{Cabrit88}.  However, this estimate is also uncertain.

Our measurements do not suffer from any of these uncertainties.  The
H$_2$ lines are optically thin and our observations encompass the most
emissive H$_2$ rotational lines, which we simply sum over rotational
states and positions to derive the total emission.  We have used shock
models to calibrate the amount of cooling accounted for by other
molecules.  In 3 sources covered by our maps, we have confirmed that
the models are accurately reproducing the observed distribution of
cooling power.  Since H$_2$ is the dominant species and a primary
coolant we are directly tracing the energy loss in the most abundant
species.  H$_2$ does not emit within the core itself and thus we have
no line of sight contamination.  By using the cooling luminosity to
trace outflow energetics our measurements are completely independent
of the outflow inclination.  As an example our momentum flux estimates
are a factor of $\sim 4-8$ above values -- uncorrected for
inclination -- derived from CO emission in same flows in NGC~1333
\citep{Hatchell07}. On the other hand, our estimates are in good
agreement the values (corrected for inclination) of \citet{Sandell01}.

Our conclusion is similar to that of \citet{Giannini01} who used ISO
observations of FIR cooling lines as a measure of the total mechanical
luminosity.  The major difference is that we use spatially resolved
observations of H$_2$ and supplement our results with those of
\citet{Giannini01} to derive the total cooling loss.  Given that we
have observed the major coolants, the primary uncertainty in the mass
loss rate derivation is the assumed shock velocity.  The expected
range of shock strengths to create the observed H$_2$ emission is
$\sim$10 -- 50 km/s, with adopted values of $\sim 20-30$ km/s
(Table~\ref{tab:shock-model-parameters}).  We therefore estimate that
the mass outflow rate values are accurate to within a factor of 3
(given a distance of 220 pc). 
% In conclusion, we have arguably obtained the most reliable measurement
% of outflow properties from -- and the mass infall rate onto -- young
% embedded protostars to date.
 
\subsubsection{Impact of flow on natal core}

It is well known that outflows inject significant energy and momentum
in the the surrounding cloud \citep[e.g.][]{Lada85,Arce07,Davis08}
with some suggestions that these flows drive supersonic motions
\citep{MacLow04}.  It is also possible that the outflow is the key
player in the disruption of the natal core
\citep[e.g.][]{Tafalla97,Fuente02}. Our data offer a new opportunity
to explore the question of the outflow impact on the surrounding
material in the specific case of low mass star formation.

The total momentum injected by outflows into a core is given by $P =
\dot{P} \, \tau_{dyn} \sim 0.3$ M$_\odot$ km s$^{-1}$, where $\dot{P}
= \dot{M}_{w} \, v_{s}$ and $\tau_{dyn}$ is the outflow dynamical
timescale. Using $\tau_{dyn}$ values from \citet{Knee00}, we find $P$
values ranging from 0.1 to 0.4 M$_\odot$~km~s$^{-1}$ (see
Table~\ref{tab:outflow-properties}). If a similar level of outflow
activity persists during the lifetime of the embedded phase of $\sim 5
\times 10^5$ yrs \citep{Evans08} then the total momentum over this
phase is $\sim 4 - 20$ M$_\odot$~km~s$^{-1}$. There is some evidence
in the literature of a decline in the outflow momentum flux during the
transition from Class 0 to Class I \citep{Bontemps96} However,
\citet{Hatchell08}, with a homogeneous sample of sources in Perseus,
do not confirm this result finding similar momentum fluxes for sources
with comparable luminosities and masses.

Based on these estimates when the entrained outflowing material slows
down to 1 km s$^{-1}$ then it will have swept up $4 - 20$ M$_\odot$
of material.  Assuming a typical core mass of $\sim 5$ M$_\odot$ then
the outflow is clearly capable of destroying the core during the
lifetime of the embedded phase.  Of course, the flow will only
disperse material in its path and the typical outflow cone angle in
these sources is $\sim 60^{\circ}$ \citep{Jorgensen07b}.  However,
observations suggest that the outflow cone angle increases with age
\citep{Velusamy98,Arce06}.  Estimates of core masses vary by nearly an
order of magnitude in the literature
\citep{Lefloch98,Sandell01,Walsh07} so definitive comparisons within
our small sample are difficult.  Nonetheless these data clearly
suggest that the outflow could be the main mechanism for core
dispersal.

To support our assertion that the transfer of outflow momentum to the
core is a main mechanism for core dispersal we can also compare the
energetics.  \citet{Tafalla97} demonstrated that the outflow from a
massive star (Mon R2) exceeds or is comparable to the gravitational
binding energy of its core.  Thus the Mon R2 outflow has likely
sculpted that object.  However massive star outflows are more
energetic than the low mass objects sampled in our maps and this
result may not scale to low mass systems.  The average total energy
generated by the current generation of outflows in the center of
NGC~1333 is $\langle L_{H_2}/f_c \times \tau_{dyn} \rangle = 2 \times
10^{43}$ ergs.  The gravitational binding energy of a sphere is
$\alpha \, G M^2/R$ with $\alpha = 1$ for cores with a $r^{-2}$ density
profile \citep{MacLaren88}, as expected for these embedded protostars.
For a typical core mass of $\sim 5$ M$_{\odot}$ and radius of 0.03 pc
the binding energy is $\sim 4 \times 10^{43}$ ergs.  If we assume that
a similar level of activity exists over the lifetime of the embedded phase
 then the outflow energetics are equal to
or even exceed the gravitational energy of the core.  In sum, both the
flow energetics and momentum suggest that outflows are the primary
method for core dispersal. Molecular gas observations show that
NGC~1333 cores discussed here are embedded in a much larger but lower
density cloud \citep[e.g.][]{Pineda08}. As typical for GMCs, the cloud
binding energy is much greater than those of each individual
cores. Thus the flows which disrupt the cores are not major players in
cloud disruption as the cloud momentum and energy far exceed that
provided by the current generation of flows over their lifetime.

\section{Conclusions}
\label{sec:conclusions}

We have presented Spitzer-IRS maps of seven pure rotational
$\mathrm{H_2}$ lines in the NGC~1333 star forming region. These
observations cover a region roughly $20\arcmin \times 13\arcmin$ that
encompass a number of YSOs and outflows. We have analysed these
observations using the rotational diagram technique in order to derive
the excitation temperature and the $\mathrm{H_2}$ ortho-to-para
ratio. Furthermore, these observations have been compared with the
predictions of a planar stationary shock model. Finally, we have used
the $\mathrm{H_2}$ total luminosity to estimate the mass outflow
rates, the mass infall rate onto the central objects, and the momentum
and energy injected into the cores and the cloud. Our main conclusions
are as follow.

\begin{itemize}

\item H$_{2}$ line emission is detected along several outflows in
  NGC~1333. In particular, we detect line emission along both lobes of
  the HH~7-11 outflow, along the south-west lobe of the outflow driven
  by IRAS~4A, along the north, south and east lobes of the IRAS~2
  outflow, as well as around IRAS~7. In addition, we detect some faint
  emission south of SVS~13, which is associated with the outflow
  originating from SVS~13B. Diffuse emission is detected north of this
  source, and might be associated with this outflow as well.

\item A rotational diagram analysis on several aperture averaged
  spectra indicate the presence of at least two gas components towards
  each position. The warm component has a rotational temperature
  between $\sim$ 300 and 600 K and an \emph{opr} between $\sim$ 0.3
  and 0.7, while the hot component has a rotational temperature
  between $\sim$ 1000 and 1500 K and an \emph{opr} between $\sim$ 1.9
  and 2.2.

\item Comparison of the line fluxes measured towards these positions
  with the predictions of a planar stationary shock model indicate the
  presence of slow and fast shock components. A good fit of the
  observation is obtained for an initial \emph{opr} of 0.01 (except in
  one source where the data are better fit with an initial \emph{opr}
  of 1), a pre-shock density of $10^{4}$, and a shock velocity of
  $\sim$20 km/s and $\sim$45 km/s the slow and fast shock components,
  respectively. This indicates, in agreement with earlier studies,
  that H$_{2}$ is mostly in para form in dense molecular clouds
  (i.e. $opr \lesssim 1$). This is consistent with the predictions of
  chemical models, provided that the cloud lifetime exeed a few
  million years.

\item Maps of the rotational temperature and the \emph{opr} in warm
  components show some important variations of these quantities across
  the mapped region. We confirm the presence of gas with a relatively
  high rotational temperature and low \emph{opr} at the tip of the
  HH~7-11 outflow.  This gas is probably ``fresh gas'' with a low
  initial \emph{opr} that has been heated-up by the passage of a
  shock, but whose \emph{opr} hasn't had time to increase
  significantly. The same feature is observed on both lobes of the
  IRAS~7 flow, although less clearly. Spatial averages of the
  \emph{opr} and the rotational temperature along the flow support
  this hypothesis. Interestingly, no spatial correlation between the
  two quantities is found.

\item Comparison between the total H$_{2}$ luminosity and the
  luminosity of other lines observed by ISO in HH~7-11 and IRAS~4
  indicates that between a fourth and half of the total cooling in
  these outflows occurs through the H$_{2}$ lines. Therefore H$_{2}$
  lines can be used to estimate the kinetic energy injected into the
  natal cloud by these flows, as well as outflow mass loss rate which
  are found to be $0.6-2 \times 10^{-6}$ M$_{\odot}$~yr$^{-1}$. The
  latter quantity also places indirect constraints on the mass
  accretion rate onto the protostar: dynamical models predict that the
  outflow mass loss rate is typically 10 times lower than the
  accretion rate. Using the dynamical timescale obtained from
  millimeter observations, we derived the total momentum injected in
  the clouds by the outflows, and compared it to the core and cloud
  binding energy. We found that outflows have the potential to disrupt
  individual cores, but are probably not major players in cloud
  disruption.
\end{itemize}

\acknowledgments This work is based on observations made with the
Spitzer Space Telescope, which is operated by the Jet Propulsion
Laboratory, California Institute of Technology under a contract with
NASA. Support for this work was provided by NASA through an award
issued by JPL/Caltech for program \#20378.

{\it Facilities:} \facility{Spitzer (IRS)}

\appendix

\section{Variation of the opr with the temperature}
\label{sec:variation-opr-with}
 
As mentioned in \S~\ref{sec:h_2-column-density}, some important
spatial variations in both the \emph{opr} and the rotational
temperature are observed in NGC~1333. In particular, we observe, at a
tip of the HH~7-11 outflow, a bow shock-like region with a low
\emph{opr} ($\sim 0.5$) and high rotational temperature ($\sim 650$ K;
see
Fig. \ref{fig:h2-rotational-temperature-opr-map-svs13-iras4}). This
low \emph{opr} region wraps nicely around the H$_{2}$ $S(1)$
rotational line emission, and suggests that it corresponds to gas
which has been recently heated-up by the passage of a shock, but whose
\emph{opr} has not had time to reach the equilibrium
value. Interestingly, this region also corresponds to the part of the
flow where the rotational temperature is the highest. Around IRAS~7,
we also observe low \emph{opr} regions around the H$_{2}$ $S(1)$
emission contours (see
Fig. \ref{fig:h2-rotational-temperature-opr-map-iras2-iras7}), but
variations from one pixel to the other are important and it is less
clear if this region correspond to high temperature gas, like in
HH~7-11.

It is interesting to compare the variations of the \emph{opr} and the
rotational temperature as a function of the position of the flow in a
more quantitative fashion. In order to do this, we have computed the
average of the \emph{opr} and rotational temperature along the
outflows that show some degree of symmetry, and where both quantities
were measured in a sufficient number of pixels. In order to
``capture'' the spatial variations, we have defined a series of
ellipses of the same eccentricity, oriented along the flow and with an
apex that is centered on the outflow driving source. The eccentricity
of the ellipses has been chosen in order to match the shape of each
outflow lobe as observed in the H$_{2}$ $S(1)$ emission. We have then
computed the average of the \emph{opr} ratio and the rotational
temperature in the region between two consecutive ellipses (that is
two ellipses with different major axes). On
Fig. \ref{fig:h2-rotational-temperature-opr-map-svs13-iras4} and
\ref{fig:h2-rotational-temperature-opr-map-iras2-iras7}, we show how
those ellipses where defined for HH~7-11, IRAS~7 (east and west lobes)
and IRAS~2 (south lobe). We have not computed the average for the
other outflow lobes, because the \emph{opr} ratio and the rotational
temperature had greater uncertainties.

Fig. \ref{fig:average_opr_trot} shows the average \emph{opr} and
rotational temperature, as a function of the ellipse semi-major
axis. Along the HH~7-11 outflow, we see a sharp drop in the average
\emph{opr} ratio 35\arcsec\, from SVS~13. Interestingly, the
temperature does not seem to be correlated with the \emph{opr}. It
increases slightly 0 and 35\arcsec\, (although at a 1$\sigma$ level),
and then starts to decrease. The highest temperature is reached where
the \emph{opr} ratio is minimal. Presumably, the hottest point
corresponds to the shock front. In the post-shock gas, the para to
ortho conversion has increased the \emph{opr}, but the conversion is
incomplete and the ratio does not reach the LTE value. In the
pre-shock gas the \emph{opr} has also probably increased with respect
to its initial value, but has not had time to reach the same value
that we see in the post-shock gas (i.e. $\sim$2). Note that for a
pre-shock density of 10$^4$ cm$^{-3}$, and a shock velocity of 20
km/s, the model predicts a shock front width -- defined as the region
of the shock where the temperature is greater than 1000 -- of $\sim
10^{3}$ AU, i.e. about 5\arcsec\, at the distance of NGC~1333.

The interpretation of the \emph{opr} and rotational temperature along
the other flows is less straightforward. We see an increase of the
rotational temperature along the IRAS~4 lobe, but the maximum
temperature is reached at the tip of the outflow; farther away no
H$_{2}$ emission is detected and the rotational temperature and the
\emph{opr} could not be determined. Unlike the HH~7-11, the average
\emph{opr} ratio is roughly constant, and is $\sim 2$. The temperature
in the east lobe of IRAS~7 flow seems to slightly decrease (note that
the first ellipse contains only a couple of points, so the average is
not very meaningful in this bin), while the \emph{opr} is almost
constant. On the other hand, the \emph{opr} ratio and temperature may
vary in a similar fashion that in HH~7-11. The \emph{opr} drops
sharply between 5 and 15\arcsec, while the temperature seems to
slightly increase in the same region. However, further away the
temperature appears to be roughly constant.

\bibliographystyle{apj}
\bibliography{bibliography}

\clearpage

\addtocounter{figure}{1}

\begin{figure}
  \centering \includegraphics{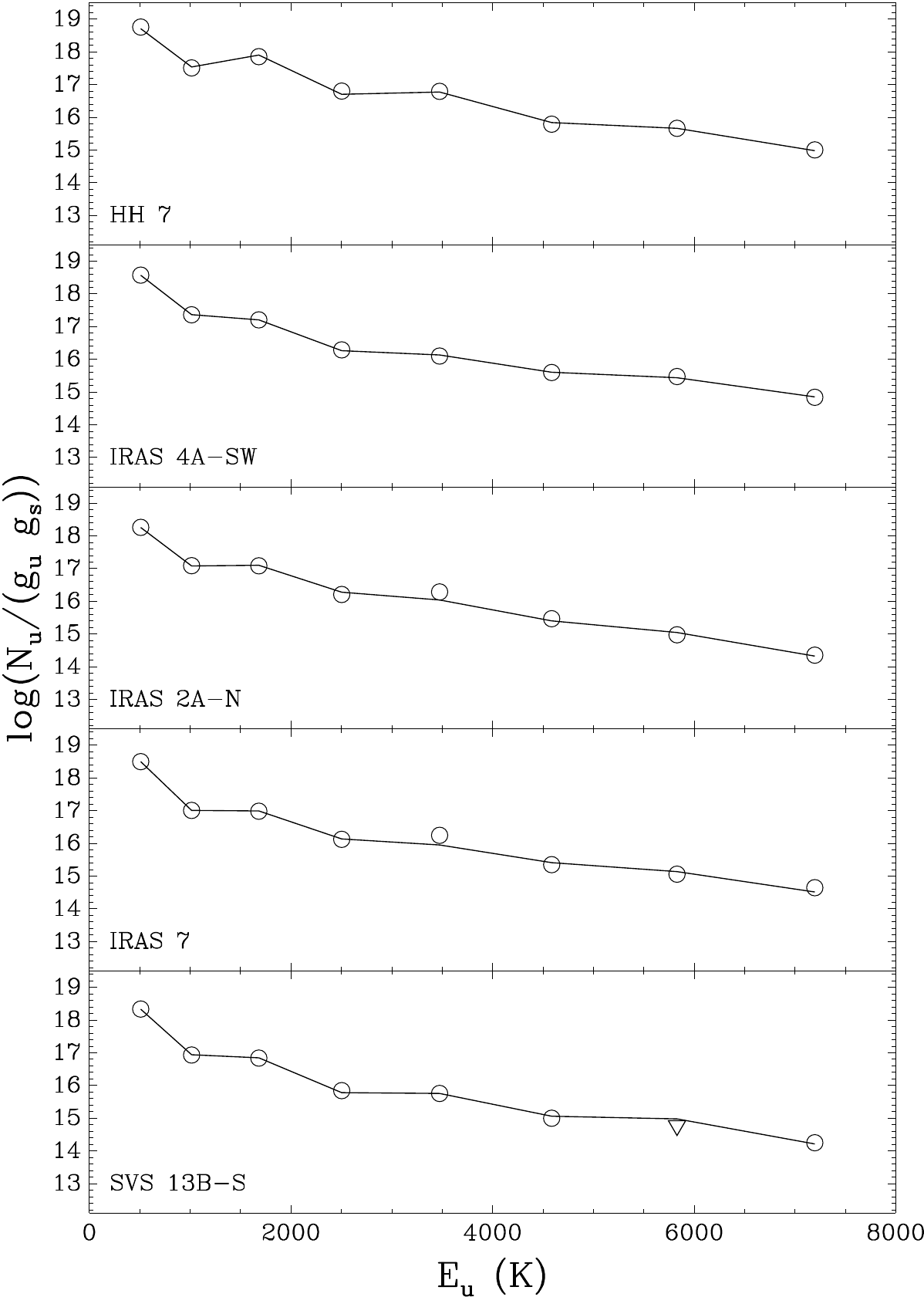}
  \caption{
    H$_{2}$ rotational diagrams obtained for 15\arcsec\ Gaussian
    apertures centered on HH 7, IRAS~4A-SW, IRAS~2A-S, IRAS~7 and
    SVS~13B-S. The black open circles correspond to the
    observations.  The black open triangle is a 1$\sigma$ upper limit. The
    black solid lines are rotational diagram obtained
    assuming that two gas components, a warm and a hot one, are present
    (see \S~\ref{sec:rotational-diagrams}). 
    \label{fig:rotational-diagram}}
\end{figure}

\clearpage

\begin{deluxetable}{lccccccc}
\tablewidth{0pt}
\rotate
\tablecaption{Line intensities}
\tablehead{
Transition & Wavelength & Module & \multicolumn{5}{c}{
Intensity (10$^{-5}$ ergs cm$^{-2}$ s$^{-1}$ sr$^{-1}$)}\\
\cline{4-8}
& ($\mu$m) &  & HH~7 & IRAS~4A-SW & IRAS~2A-N & IRAS~7 & SVS~13B-S
}
\startdata
H$_{2}$ $S(0)$ & 28.2188 & LH  &   0.45 &   0.30 &   0.14 &   0.25 &   0.17\\
H$_{2}$ $S(1)$ & 17.0348 & SH  &   2.82 &   2.02 &   1.08 &   0.89 &   0.75\\
H$_{2}$ $S(2)$ & 12.2786 & SH  &  21.28 &   4.79 &   3.66 &   2.88 &   2.05\\
%H$_{2}$ $S(2)$ & 12.2786 & L1  &  18.36 &   3.85 &   1.90 &   2.32 &   1.27\\
H$_{2}$ $S(3)$ &  9.6649 & L1  &  28.75 &   8.85 &   7.39 &   6.05 &   3.16\\
H$_{2}$ $S(4)$ &  8.0251 & L1  &  39.35 &   8.04 &  12.31 &  11.24 &   3.63\\
H$_{2}$ $S(5)$ &  6.9095 & L2  &  35.86 &  22.94 &  17.25 &  13.13 &   5.82\\
H$_{2}$ $S(6)$ &  6.1086 & L2  &  21.94 &  14.24 &   4.54 &   5.50 &$<$2.94\tablenotemark{a}\\
H$_{2}$ $S(7)$ &  5.5112 & L2  &  31.31 &  21.63 &   7.06 &  13.81 &   5.54\\
\ion{Fe}{2} $^{6}D_{7/2} - ^{6}D_{9/2}$ & 25.9882 & LH  &   0.71 &   0.49 &   0.22 &   3.34 &   0.26\\
\ion{S}{1} $^{3}P_{1} - ^{3}P_{2}$ & 25.2490 & LH  &   0.62 &   0.38 &   1.71 &   0.38 &   0.32\\
\ion{Si}{2} $^{2}P_{3/2}^{0} - ^{2}P_{1/2}^{0}$ & 34.8141 & LH  &   1.01 &   0.39 &   0.20 &   3.24 &   0.41\\
\enddata
\tablenotetext{a}{1 $\sigma$ upper limit}
\label{tab:intensities}
\end{deluxetable}

\begin{deluxetable}{l c c c c c c c}
  \tablewidth{0pt}
  \tablecaption{Rotational diagrams results}
   \tablehead{
     & \multicolumn{3}{c}{Warm component} && \multicolumn{3}{c}{Hot component}\\
     \cline{2-4} \cline{6-8}
     Source & $N(\mathrm{H_{2}})$ & $T_\mathrm{rot}$ & $opr$ &
     & $N(\mathrm{H_{2}})$ & $T_\mathrm{rot}$ & $opr$\\
     & (cm$^{-2}$) & (K) & & & (cm$^{-2}$) & (K) &
   }
  \startdata
  HH~7 & $5.9 \times 10^{19}$ & 611 & 0.37 & & 
  $6.2 \times 10^{18}$ & 1401 & 1.99\\
  IRAS~4A-SW & $5.9 \times 10^{19}$ & 342 & 0.69 & & 
  $3.4 \times 10^{18}$ & 1513 & 1.94\\
  IRAS~2A-N & $2.0 \times 10^{19}$ & 371 & 0.33 & & 
  $5.6 \times 10^{18}$ & 1055 & 2.09\\
  IRAS~7 & $4.2 \times 10^{18}$ & 300 & 0.31 & & 
  $3.3 \times 10^{18}$ & 1268 & 2.11\\
  SVS~13B-S & $3.2 \times 10^{19}$ & 312 & 0.51 & & 
  $1.5 \times 10^{18}$ & 1337 & 1.43\\
  \enddata
  \label{tab:rotational-diagram-results}
  %\tablenotetext{}{}
\end{deluxetable}

\begin{deluxetable}{l c c c c c c c c c}
  \tablewidth{0pt}
  \tablecaption{Best fit shock model parameters for aperture-averaged positions}
   \tablehead{
     & \multicolumn{4}{c}{Slow shock component} && \multicolumn{4}{c}{Fast shock component}\\
     \cline{2-5} \cline{7-10}
     Source & $n(\mathrm{H})$ & $v$ & $opr_i$ & $ff$ & 
     & $n(\mathrm{H})$ & $v$ & $opr_i$ & $ff$\\
     & (cm$^{-3}$) & (km s$^{-1}$) & & (\%) & & (cm$^{-3}$) & (km s$^{-1}$) & & (\%)
   }
  \startdata
  HH~7 & $10^{4}$ & 22 & 0.01 & 15 & & $10^{4}$ & 45 & 0.01 & 3\\
  IRAS~4A-SW & $10^{4}$ & 12 & 1 & 12 & & $10^{4}$ & 41 & 1 & 2\\
  IRAS~2A-S & $10^{4}$ & 23 & 0.01 & 4 & & $10^{4}$ & 50 & 0.01 & 1.5\\
  IRAS~7 & $10^{4}$ & 24 & 0.01 & 2.2 & & $10^{4}$ & 53 & 0.01 & 1.1\\
  SVS~13B-S & $10^{4}$ & 14 & 0.01 & 3 & & $10^{4}$ & 36 & 0.01 & 0.8\\
  \enddata
  \label{tab:shock-model-parameters}
  %\tablenotetext{}{}
\end{deluxetable}

\begin{figure}
  \centering \includegraphics[]{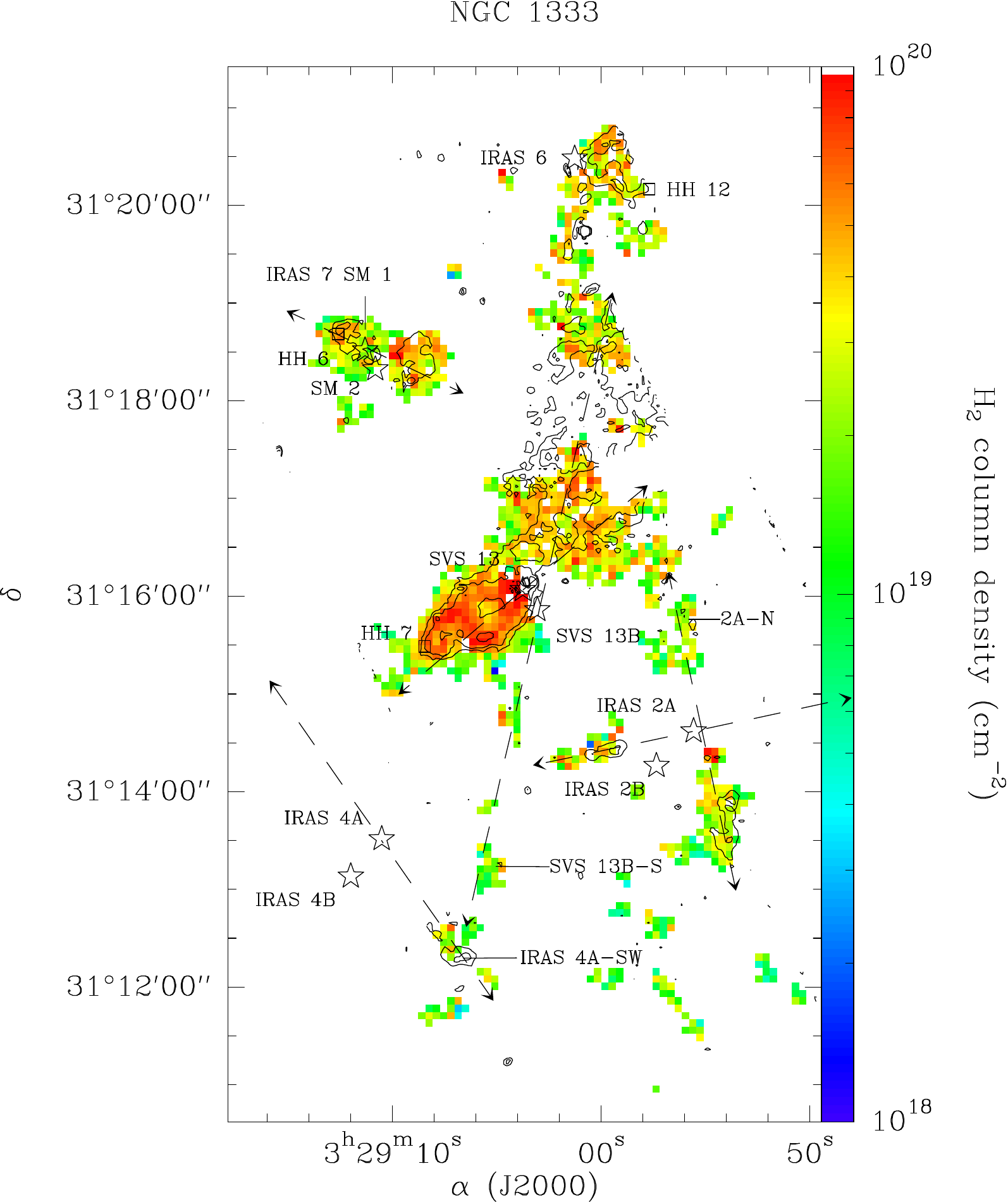}
  \caption{H$_{2}$ column density map (color points) on top of the
    H$_{2}$ $S(1)$ emission (black contours). Regions of the map that
    appear in white correspond to points where a rotational diagram
    could not be constructed because of missing data or insufficient
    signal-to-noise ratio. Contour levels are those of the $S(1)$
    emission taken from
    Fig. \ref{fig:h2-s1-map}.\label{fig:h2-column-density-map}. Pixel
    size is 5$\arcsec$.}
\end{figure}

\begin{figure}
  \centering \includegraphics[]{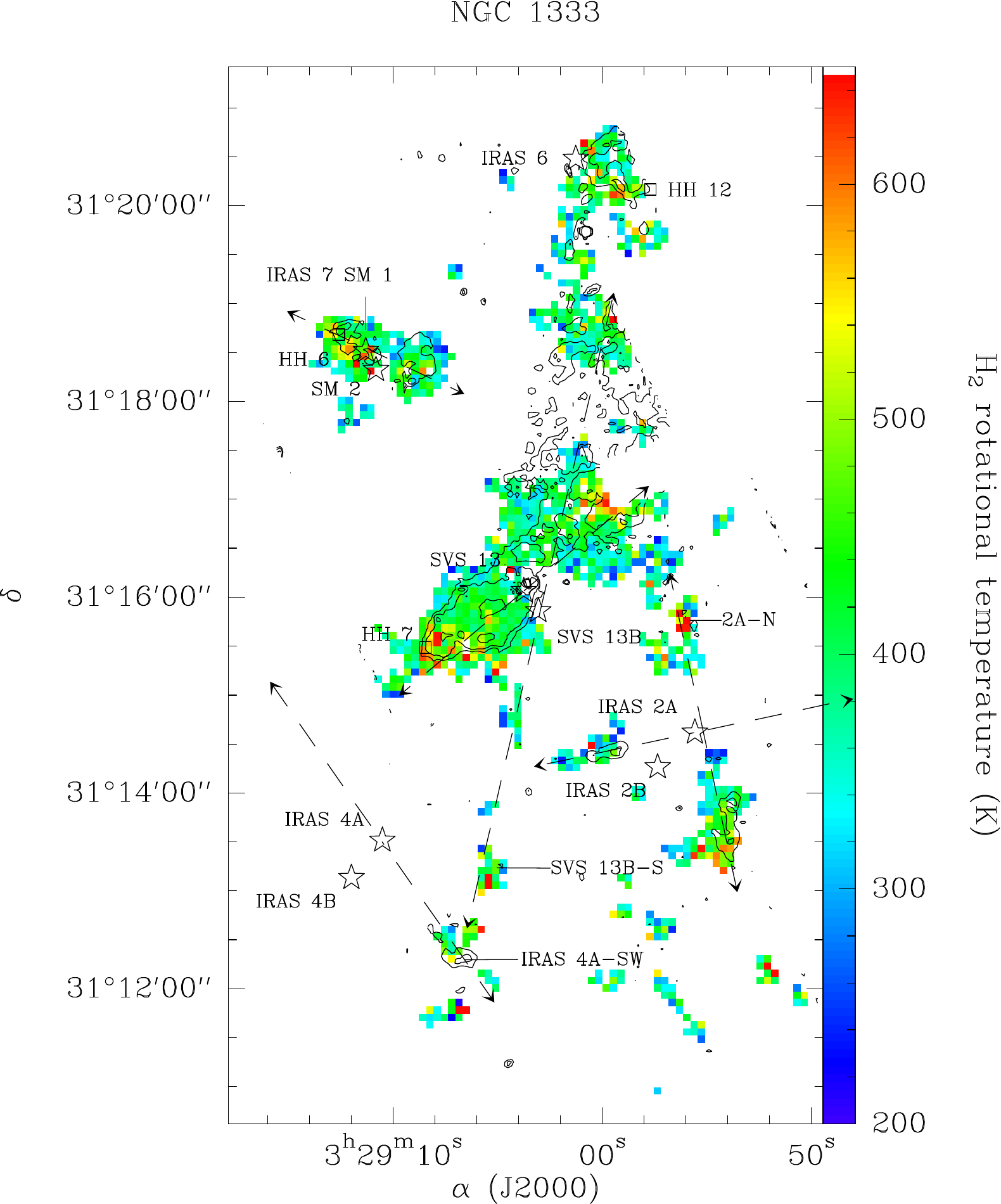}
  \caption{Same as Fig. \ref{fig:h2-column-density-map} for the
    H$_{2}$ rotational temperature.\label{fig:h2-rotational-temperature-map}}
\end{figure}

\begin{figure}
  \centering \includegraphics[]{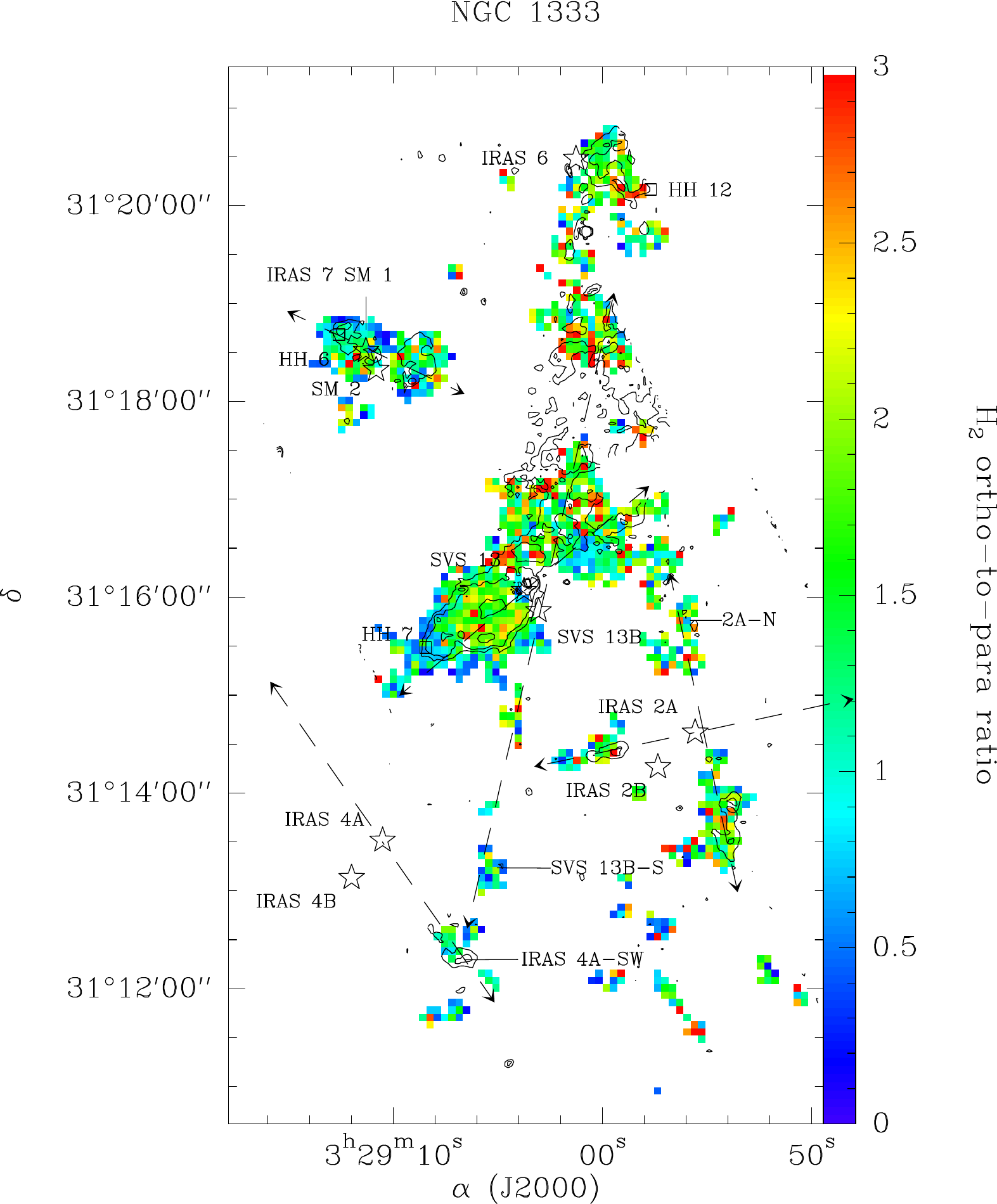}
  \caption{Same as Fig. \ref{fig:h2-column-density-map} for the
    H$_{2}$ ortho-to-para ratio.\label{fig:h2-opr-map}}
\end{figure}

\begin{figure}
  \centering \includegraphics{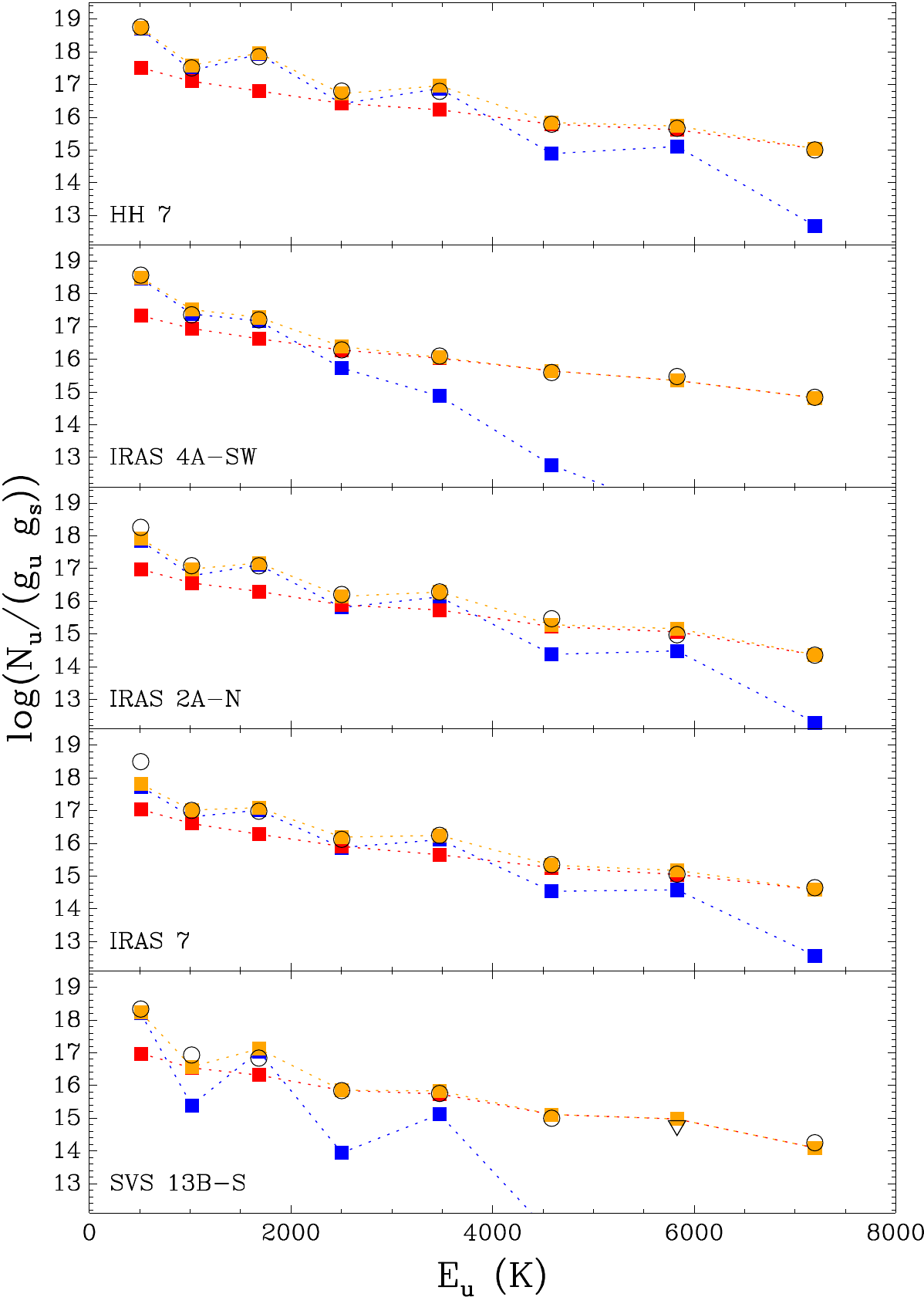}
  \caption{
    Same as in Fig. \ref{fig:rotational-diagram}, but with the two
    velocity components shock model predictions over plotted (see
    \S~\ref{sec:comp-observ-emiss}). The blue and red squares connected by
    dotted lines show the emission from the slow and fast shock
    components, respectively. The orange squares connected by dotted lines
    show the total emission from these two components. Note that some of
    the blue squares (at low energies), as well as some of the red squares
    (at high energies) are masked by the orange squares.
    \label{fig:shock-models}}
\end{figure}

\begin{figure}
  \centering \includegraphics[width=10cm]{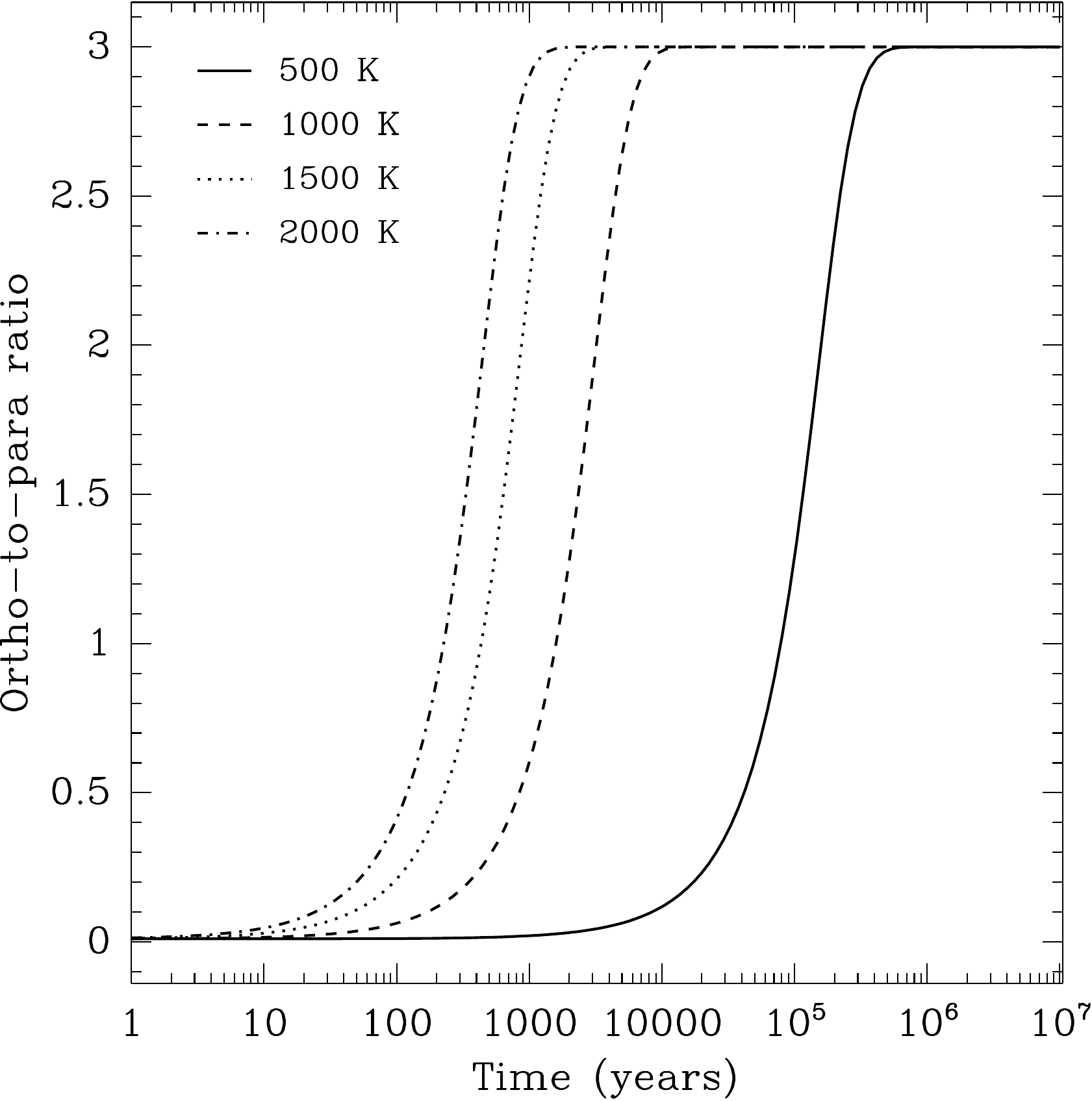}
  \caption{ H$_{2}$ ortho-to-para ratio as a function of time for
    different gas temperatures. An initial \emph{opr} of 0.01 and H
    density of 10~cm$^{-3}$ are assumed for all
    curves. \label{fig:opr-time}}
\end{figure}

\begin{deluxetable}{l c c c c c c c c}
  \tablewidth{0pt}
  \tablecaption{Outflow properties}
  \tablehead{
     Source & L$_{\rm H_2}$ & $f_c$ & $v_s$ & $\dot{M}_w$ & $\dot{P}$ &
     $\tau_{dyn}$\tablenotemark{a} & $P$\\
     & ($10^{-2}$ L$_\odot$) & & (km s$^{-1})$ & (M$_\odot$ yr$^{-1}$) &
     (M$_\odot$ yr$^{-1}$ km s$^{-1}$) & (yr) & (M$_\odot$ km s$^{-1}$)
   }
   \startdata
   HH~7-11 & 11.1 & 0.50 & 26 & 2 $\times 10^{-6}$ &  5 $\times 10^{-5}$ & 5600  & 0.29 \\
   IRAS~4A & 1.8 & 0.25 & 16 & 2 $\times 10^{-6}$  & 3 $\times 10^{-5}$  & 12000 & 0.38 \\
   IRAS~2\tablenotemark{b} & 1.5 & 0.25 & 28 & 6  $\times 10^{-7}$  & 2 $\times 10^{-5}$  & 17000 & 0.34 \\
   IRAS~7 & 1.8 & 0.25 & 34 &  5  $\times 10^{-7}$ & 2 $\times 10^{-5}$  & 5700  & 0.11 \\
   SVS~13B & 0.8 & 0.25 & 19 & 7 $\times 10^{-7}$  & 2 $\times 10^{-5}$  & 22000 & 0.44 \\
   \enddata
   \label{tab:outflow-properties}
   \tablenotetext{a}{From \citet{Knee00}}
   \tablenotetext{b}{North-south outflow only}
 \end{deluxetable}

\begin{figure}
  \begin{center}
    \begin{tabular}{cc}
      \includegraphics[width=8.0cm]{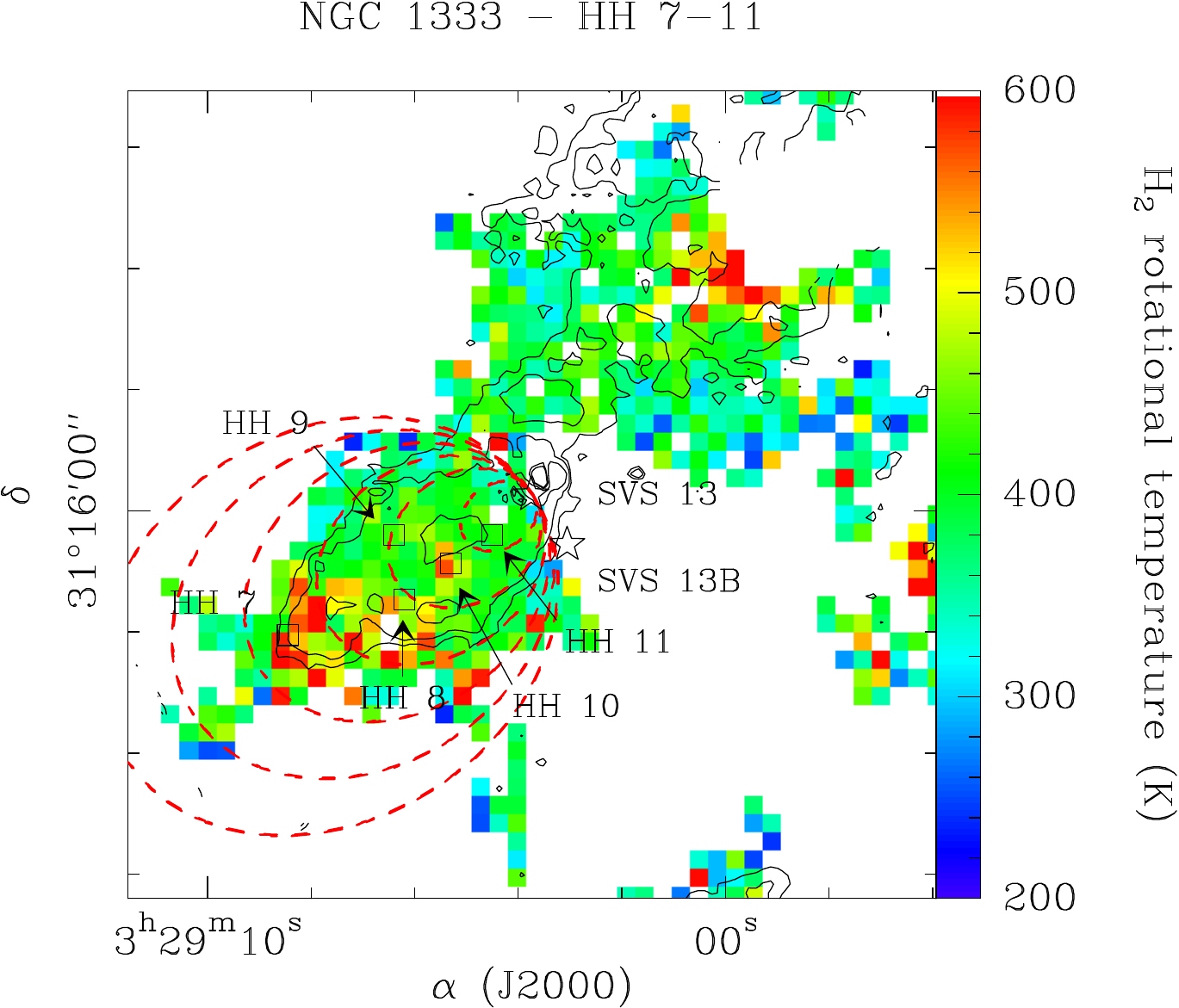} &
      \includegraphics[width=8.0cm]{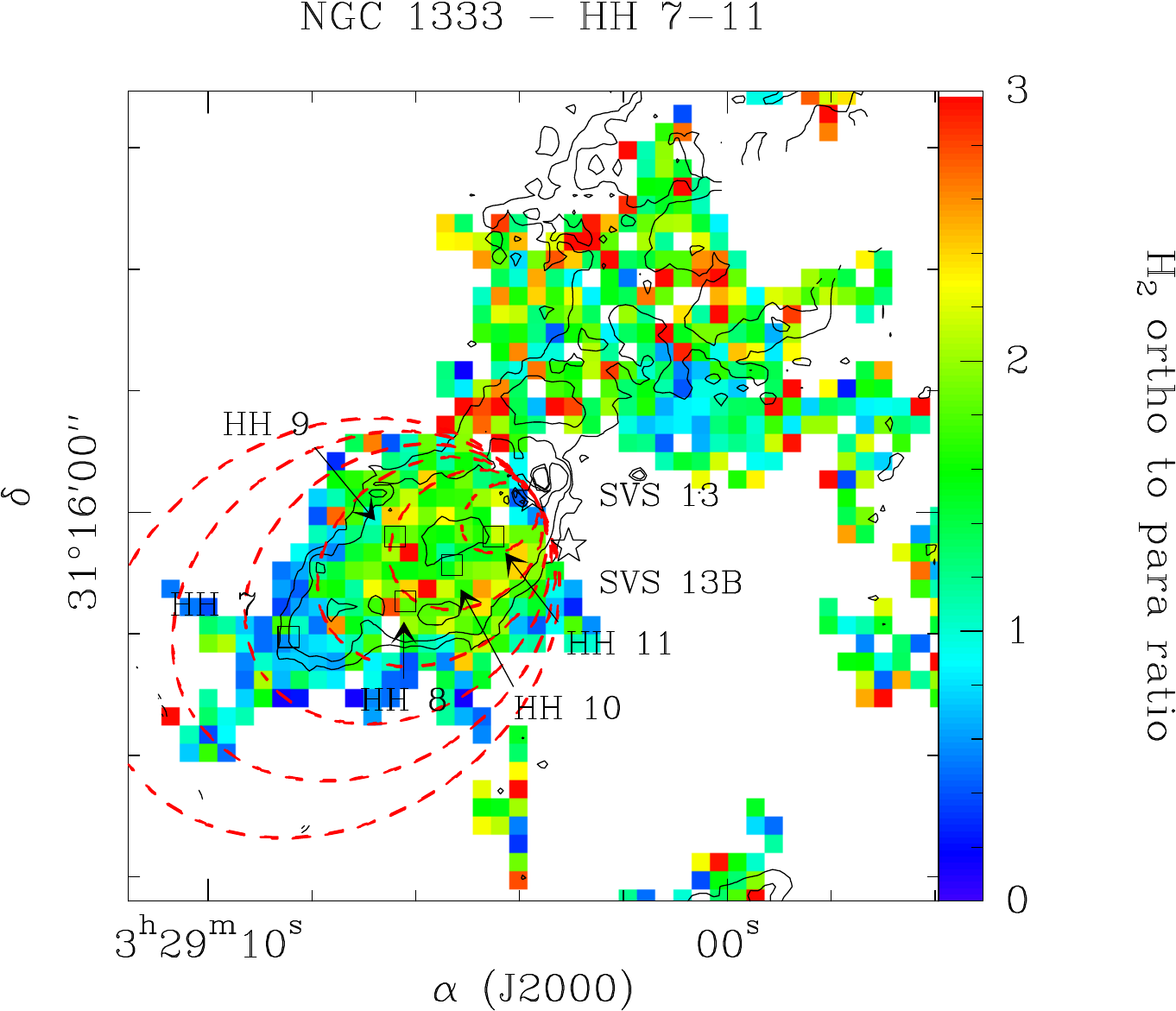}\\
      \includegraphics[width=8.0cm]{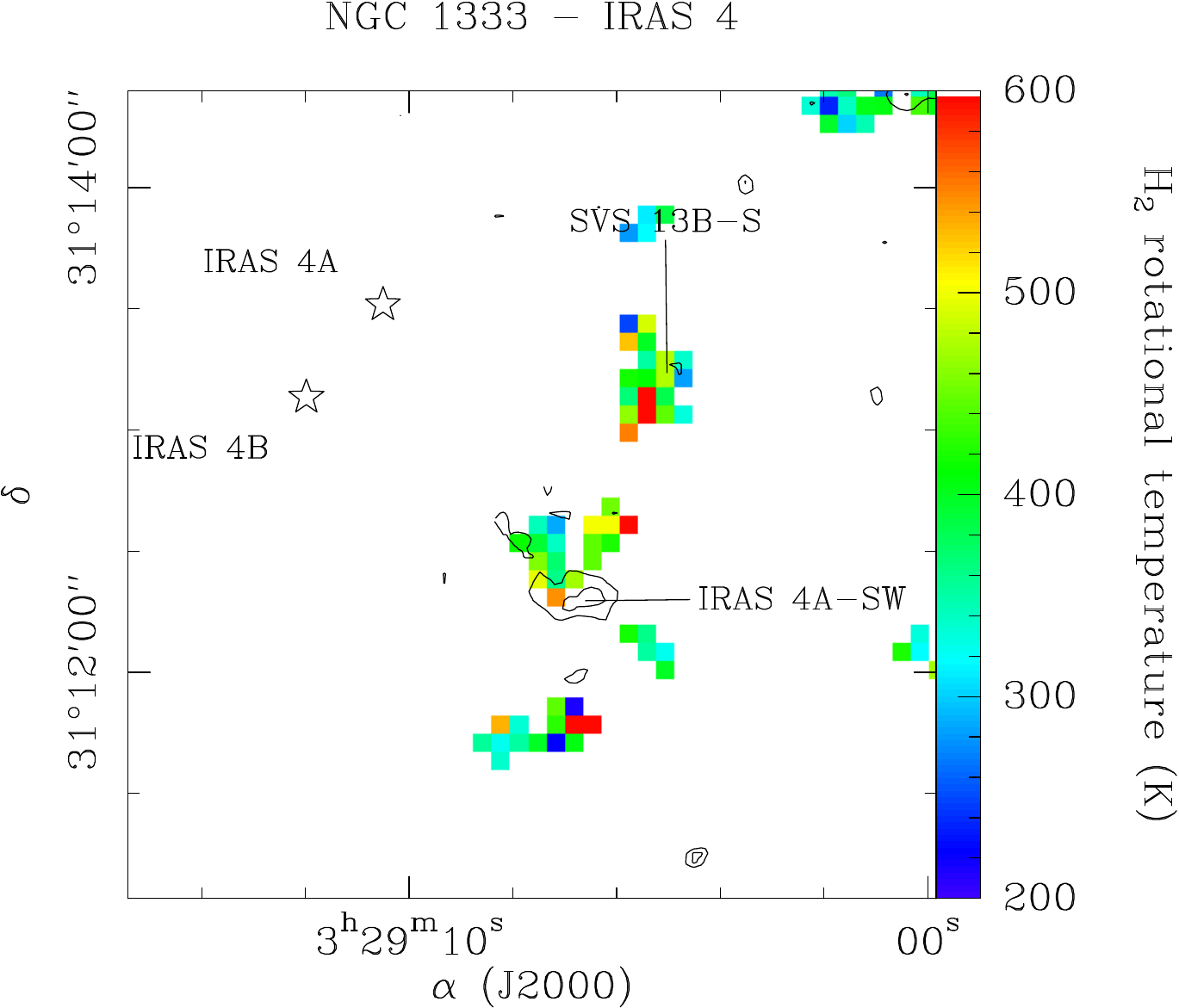} &
      \includegraphics[width=8.0cm]{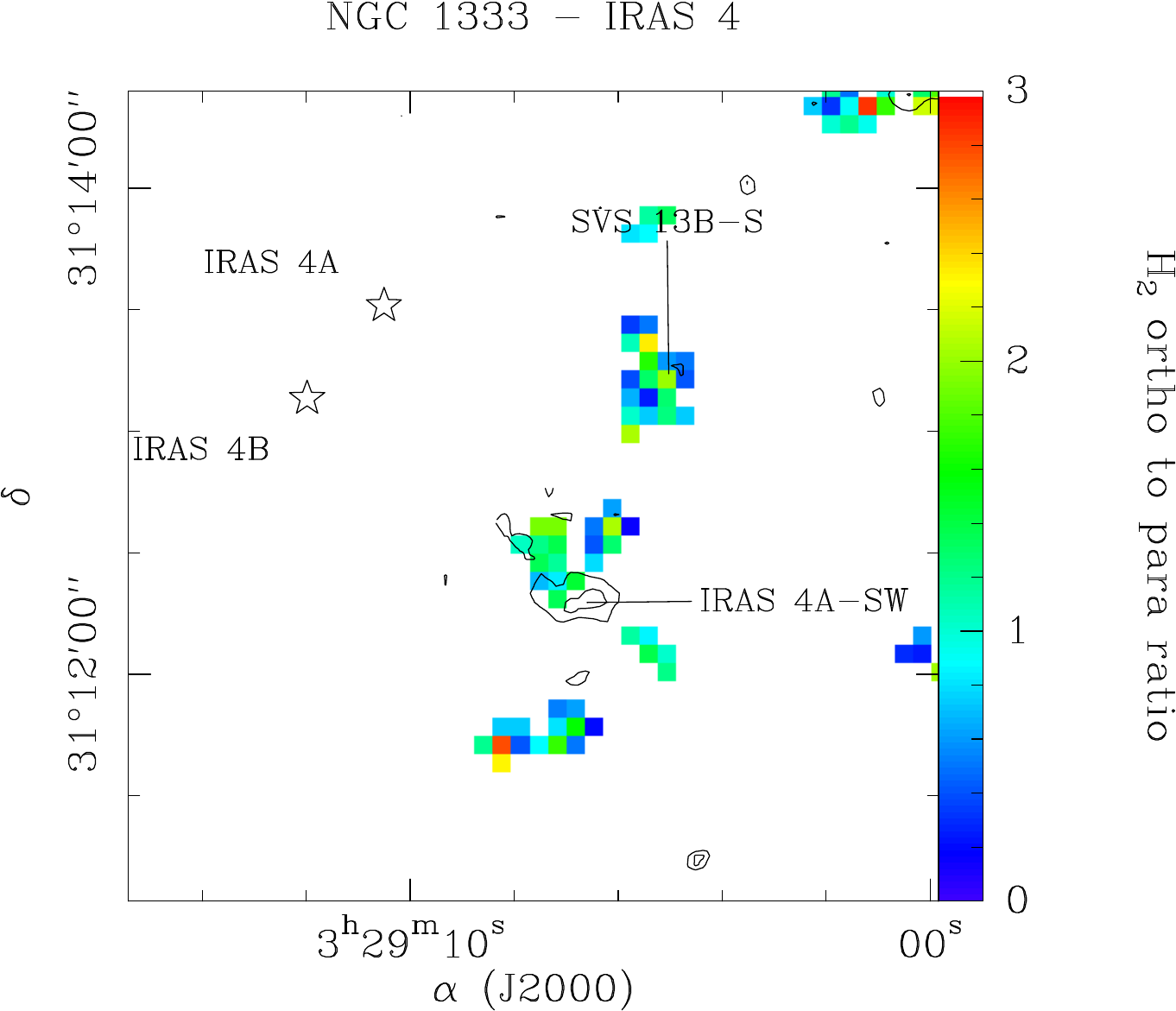}\\
    \end{tabular}
  \end{center}
  \caption{Side-by-side comparison between the rotational temperature
    and \emph{opr} around SVS~13 (upper panels) and IRAS~4 (lower
    panels). The position of several HH objects around SVS~13, from
    \cite{Walawender05a}, are
    indicated \label{fig:h2-rotational-temperature-opr-map-svs13-iras4}. The
    ellipses that were used to average the \emph{opr} and the
    rotational temperature (see \S~\ref{sec:variation-opr-with}) are
    shown.}
\end{figure}

\begin{figure}
  \begin{center}
      \begin{tabular}{cc}
        \includegraphics[width=8.0cm]{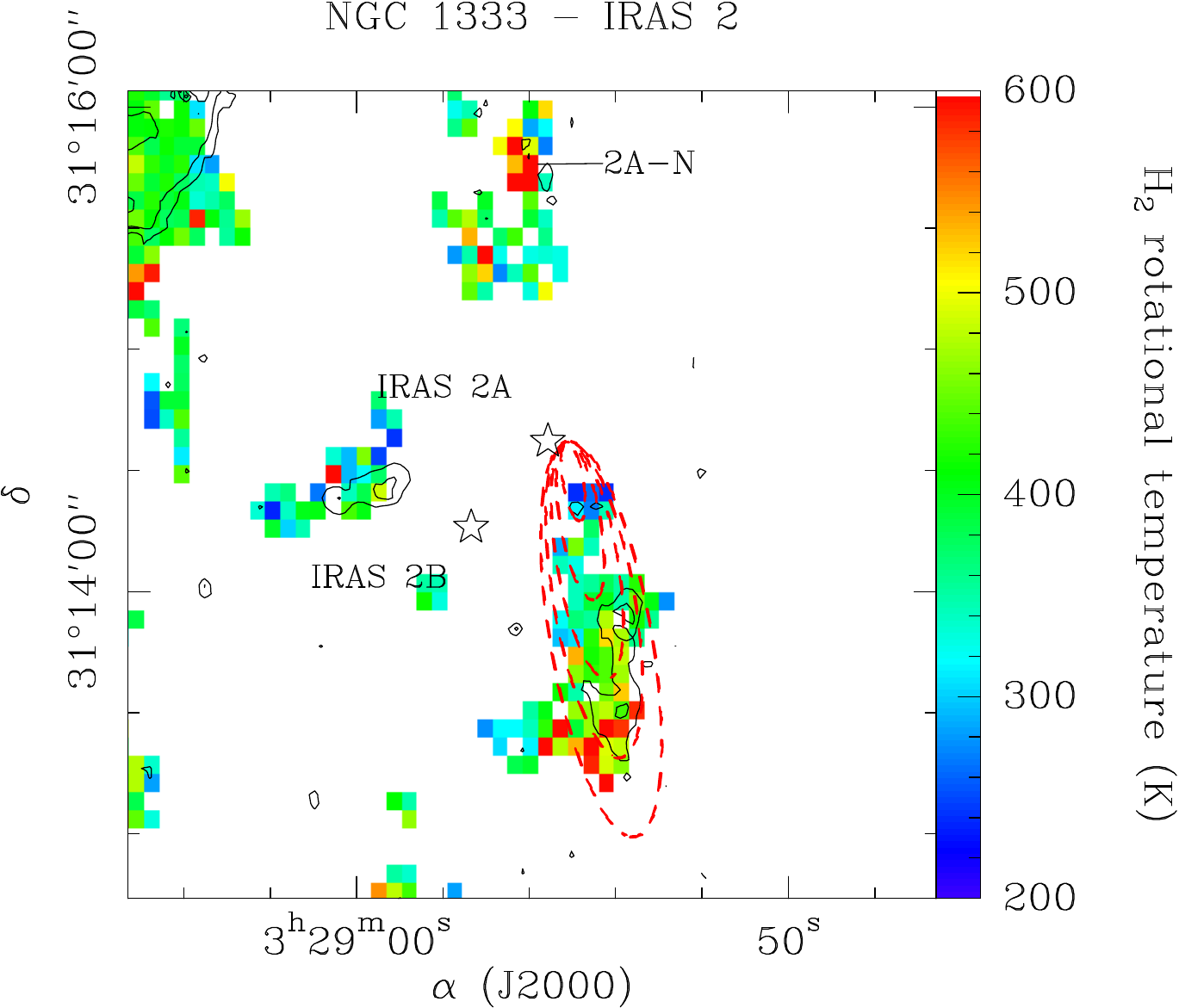} &
        \includegraphics[width=8.0cm]{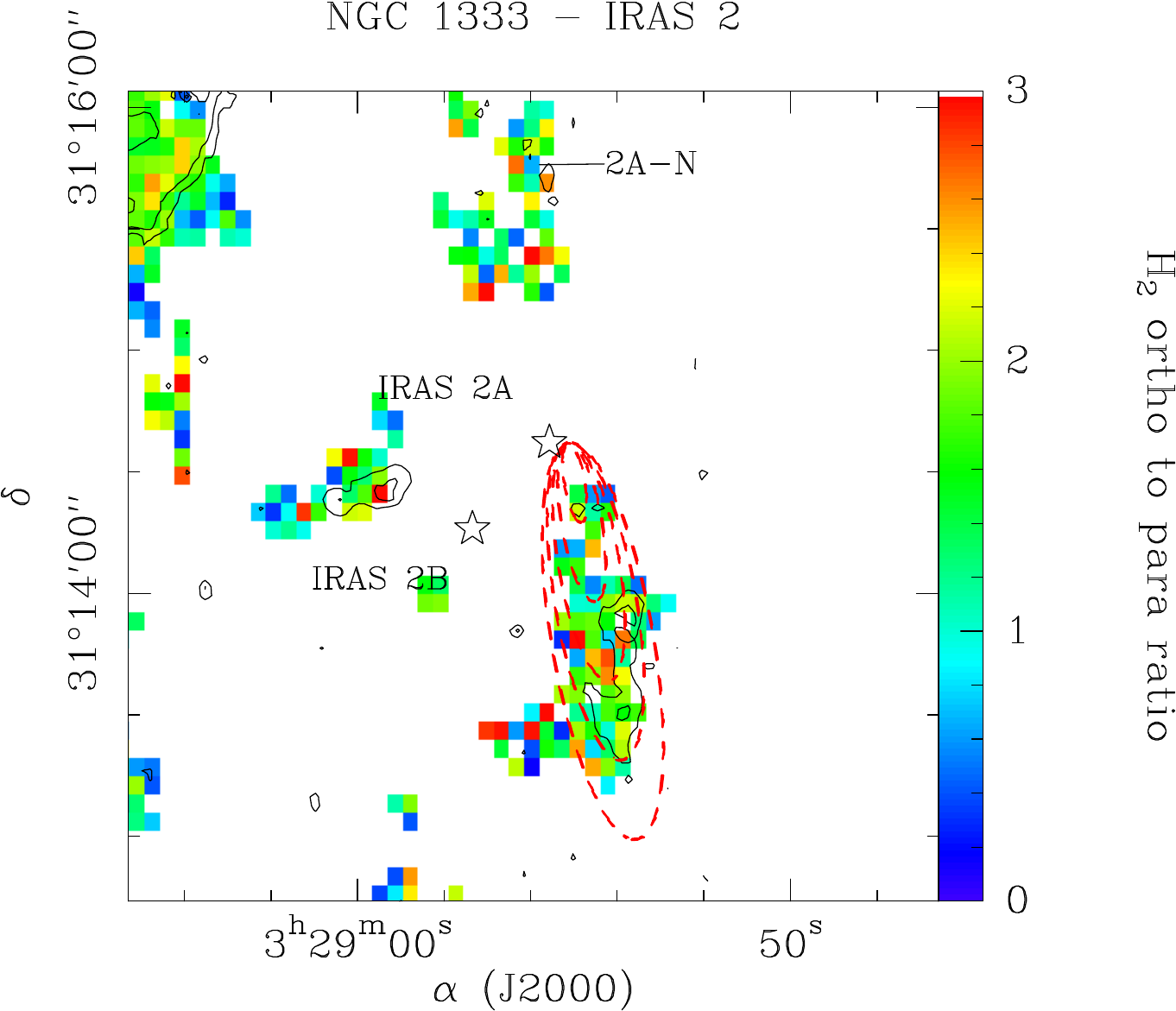}\\
        \includegraphics[width=8.0cm]{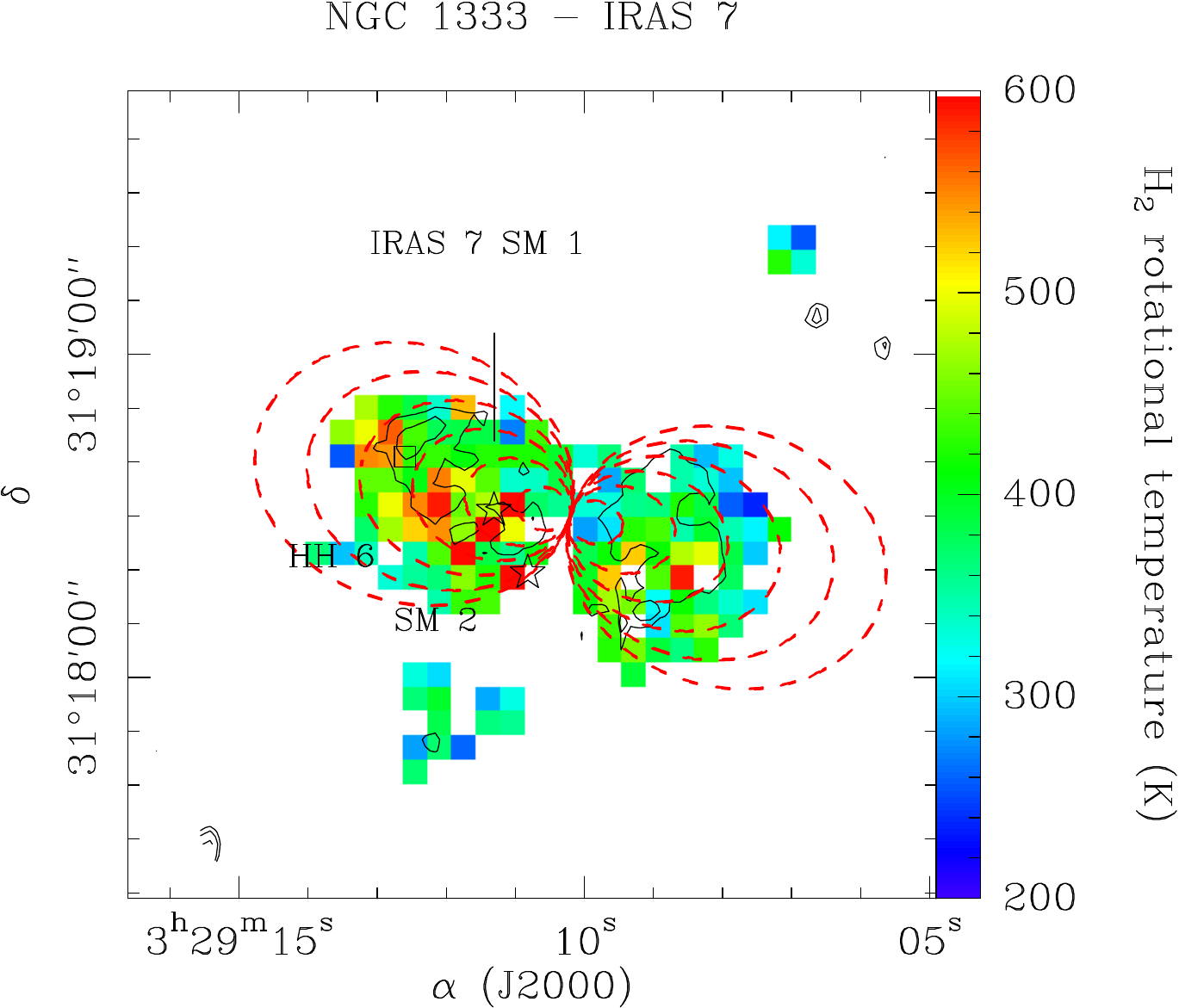} &
        \includegraphics[width=8.0cm]{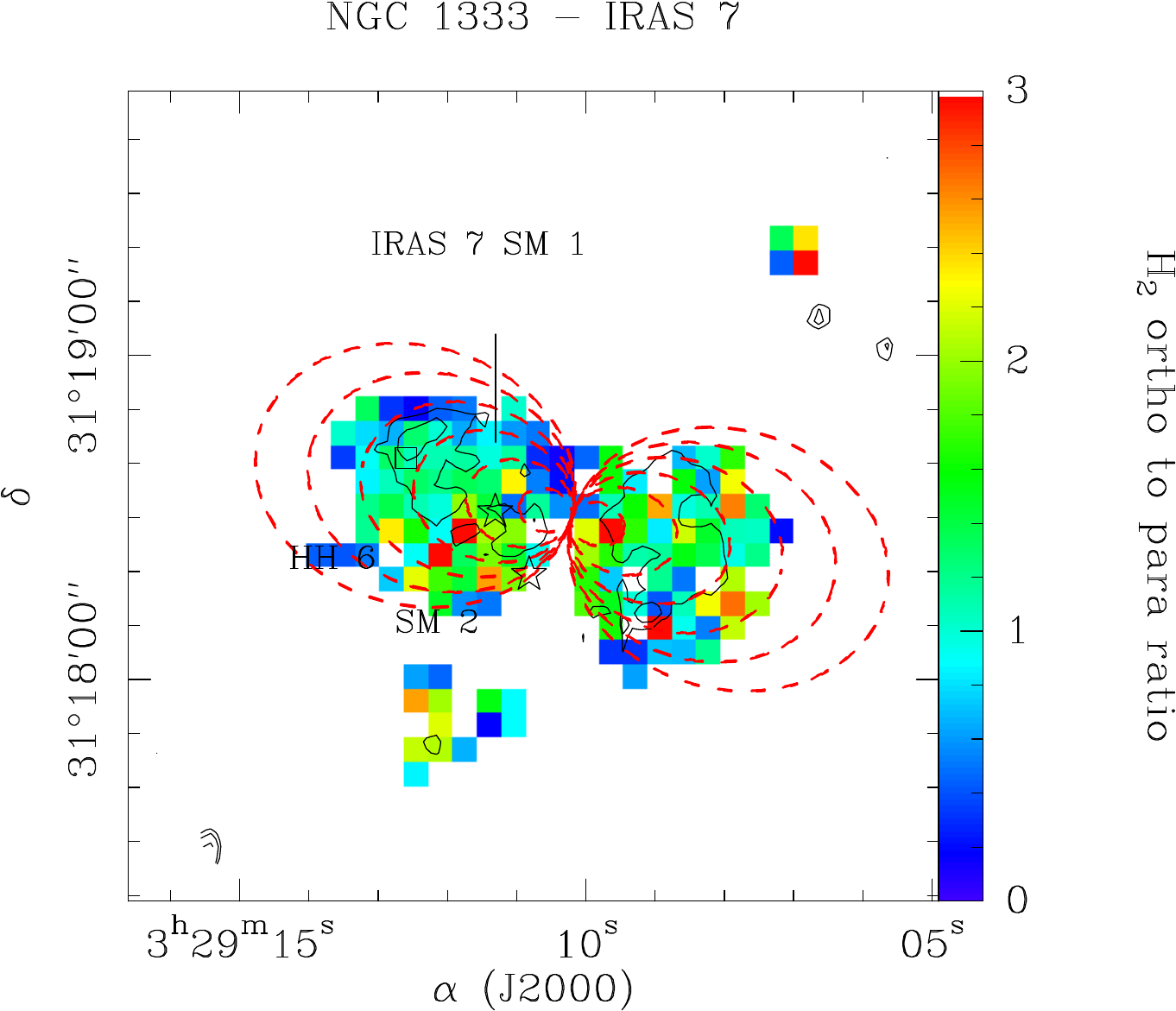}\\
      \end{tabular}
    \end{center}
    \caption{Same as
      Fig. \ref{fig:h2-rotational-temperature-opr-map-svs13-iras4} for
      the IRAS~2 (upper panels) and IRAS~6 (lower
      panels).\label{fig:h2-rotational-temperature-opr-map-iras2-iras7}}
\end{figure}

\begin{figure}
  \centering \includegraphics[]{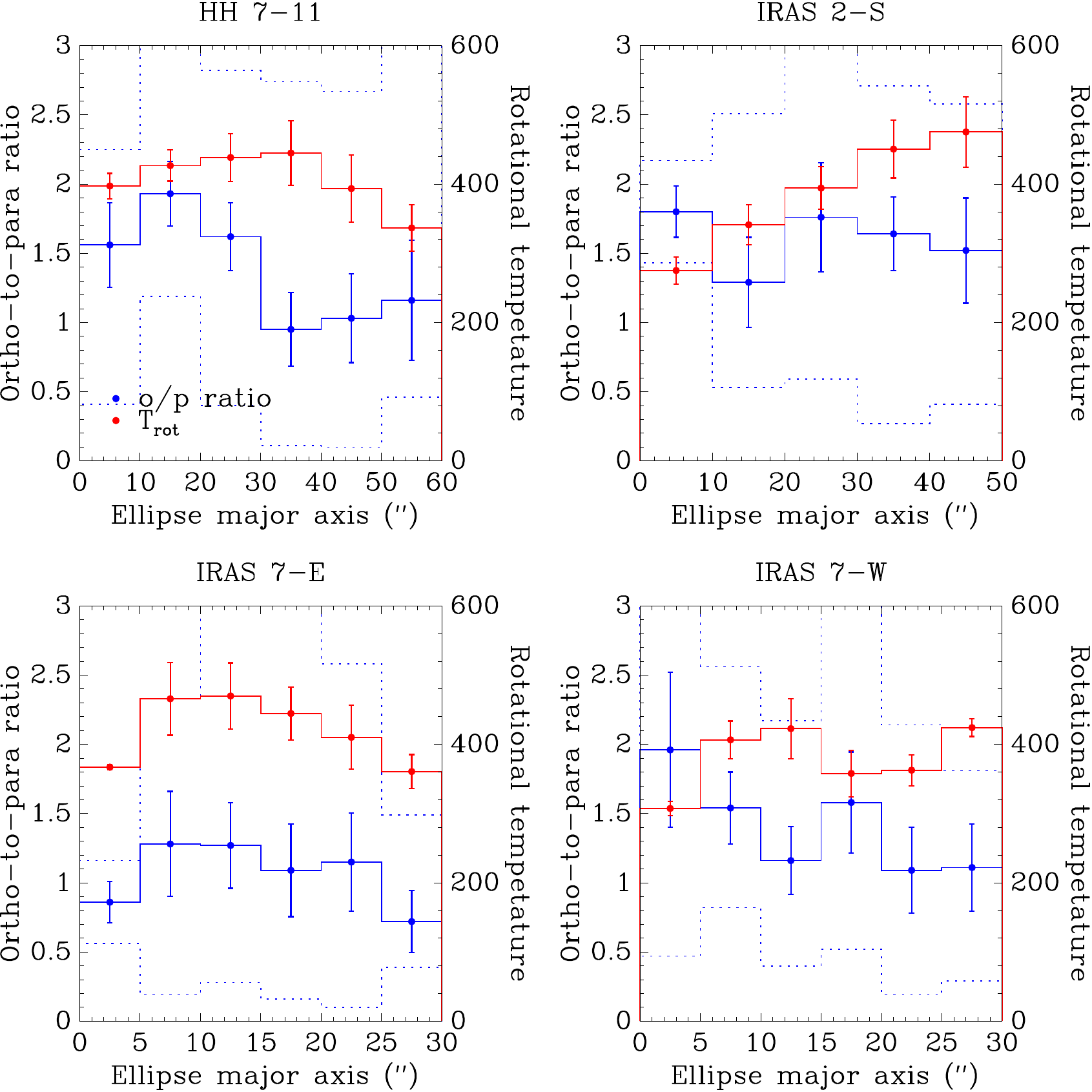}
  \caption{
    Average H$_{2}$ ortho-to-para ratio (blue histograms in solid
    lines) and rotational temperature (red histograms) between consecutive
    ellipses of the same eccentricity, as a function of the ellipse
    semi-major axis. The error bars are 1 $\sigma$ standard deviations
    computed from the variance in each bin (i.e. the region between two
    ellipses). The blue histogram in dotted lines show the maximum and
    minimum values of the \emph{opr} in each
    bin. \label{fig:average_opr_trot}
  }
\end{figure}
 
% Online figures

\begin{figure}
  \figurenum{4~d}
  \centering \includegraphics[]{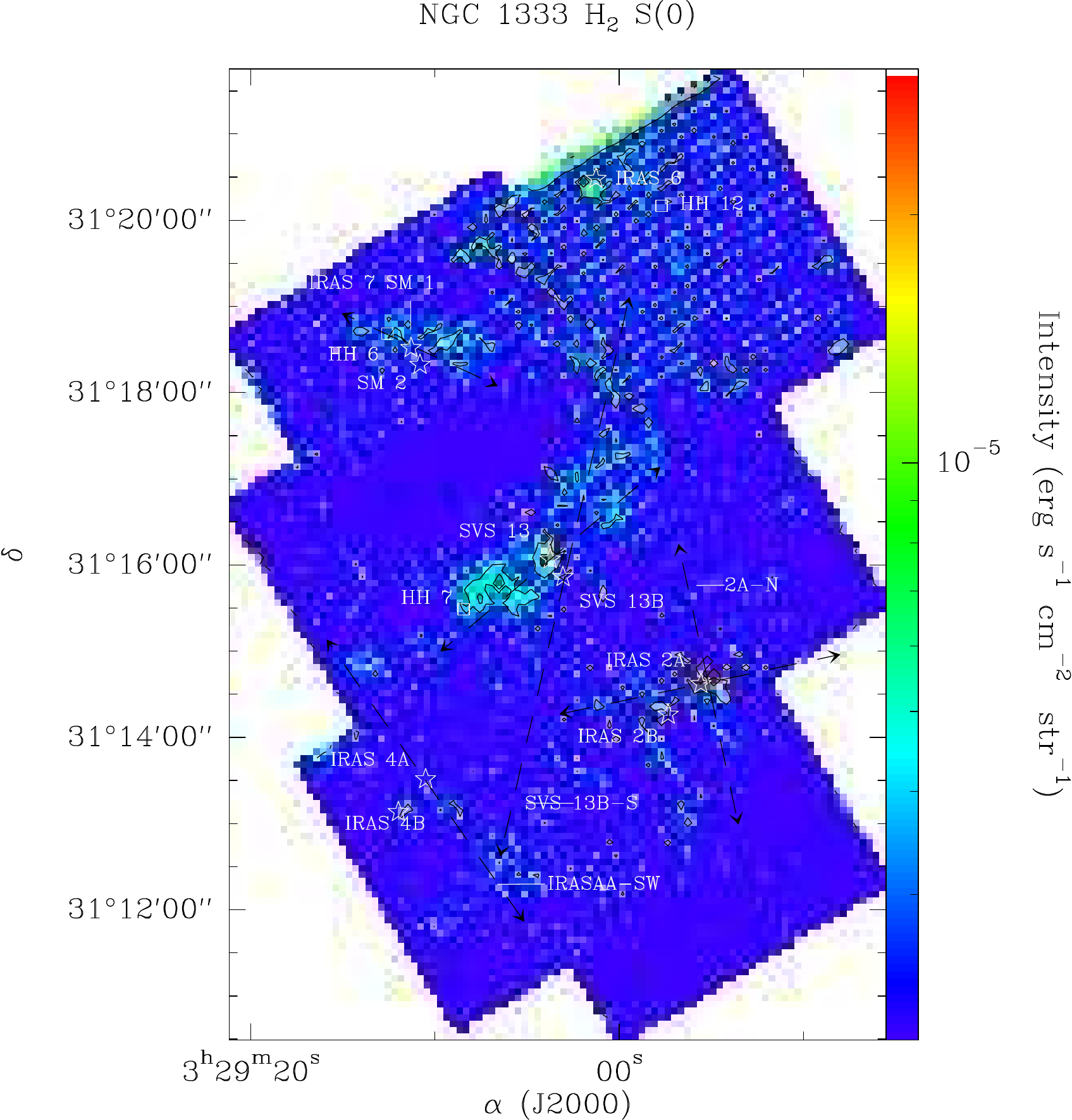}
  \caption{H$_{2}$ $S(0)$ map obtained with the LH
    module. Contour levels are drawn at 3 $\sigma$. 
    \label{fig:h2-s0-map}}
\end{figure}

\begin{figure}
  \figurenum{4~e}
  \centering \includegraphics[]{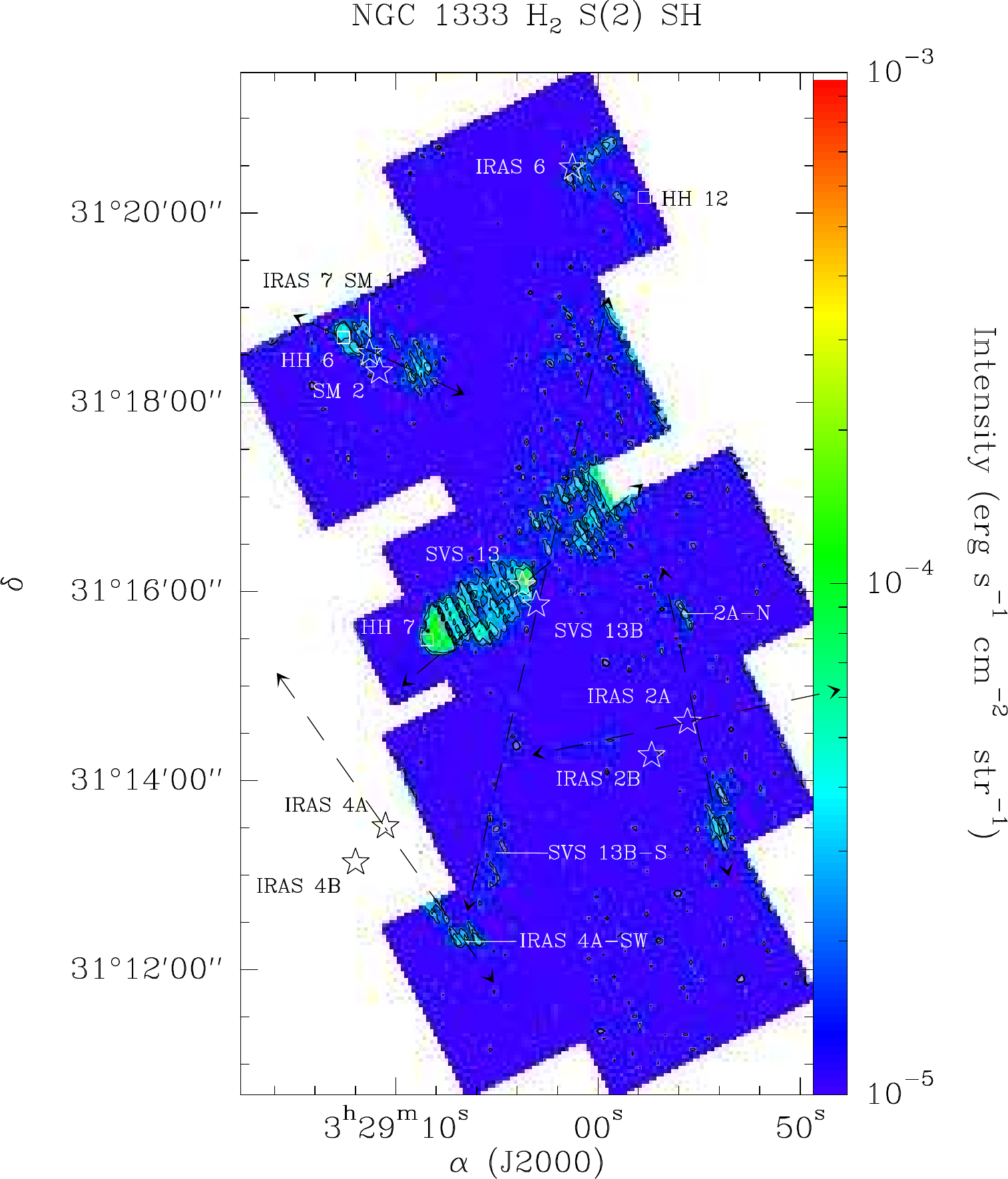}
  \caption{H$_{2}$ $S(2)$ map obtained with the SH module. Contours
    show the 3 and 5 $\sigma$ noise levels. \label{fig:h2-s2-sh-map}}
\end{figure}

\begin{figure}
  \figurenum{4~f}
  \centering \includegraphics[]{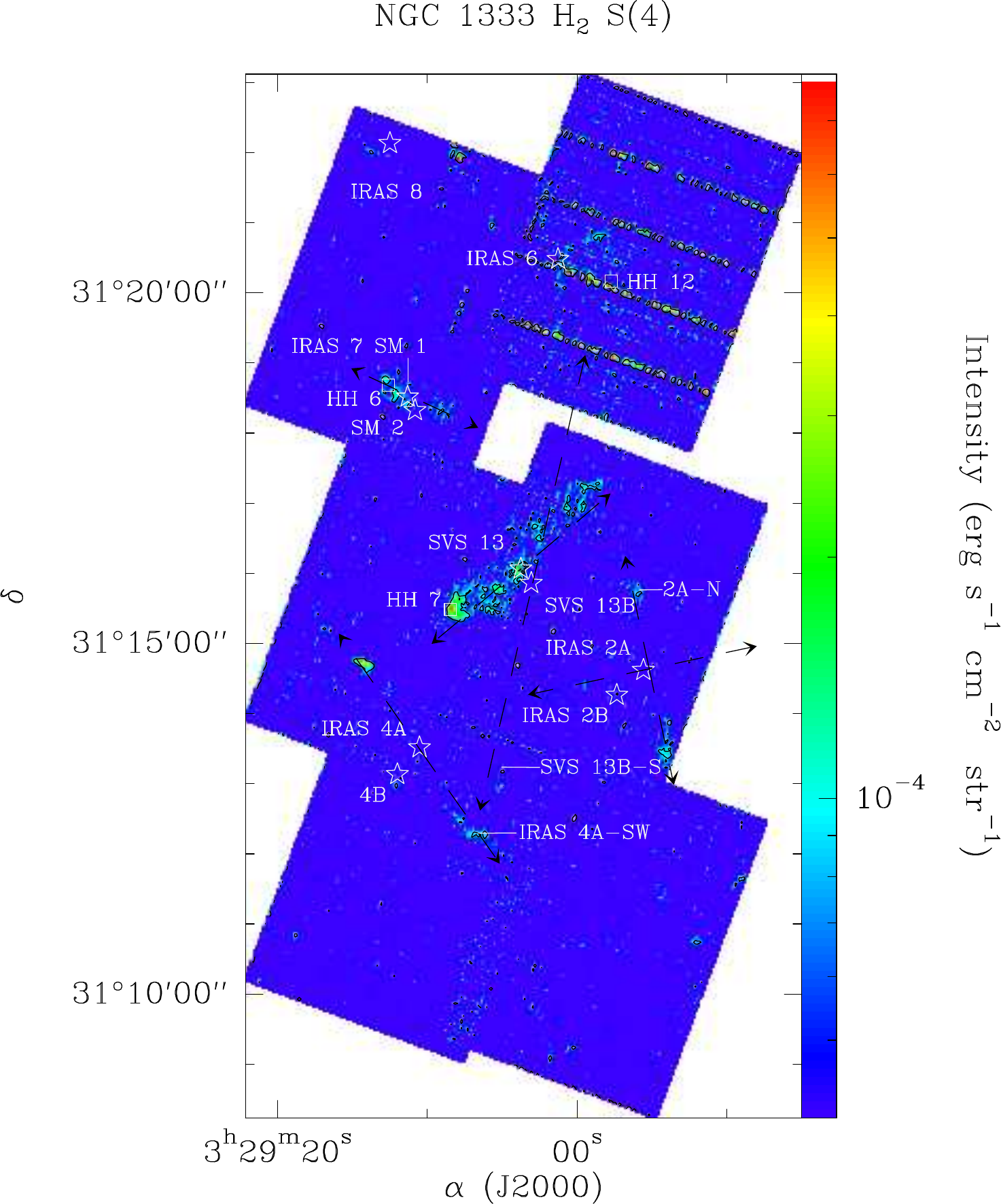}
  \caption{H$_{2}$ $S(4)$ map obtained with the SL module. Contours
    show the 3 $\sigma$ noise level. \label{fig:h2-s4-map}}
\end{figure}

\begin{figure}
  \figurenum{4~g}
  \centering \includegraphics[]{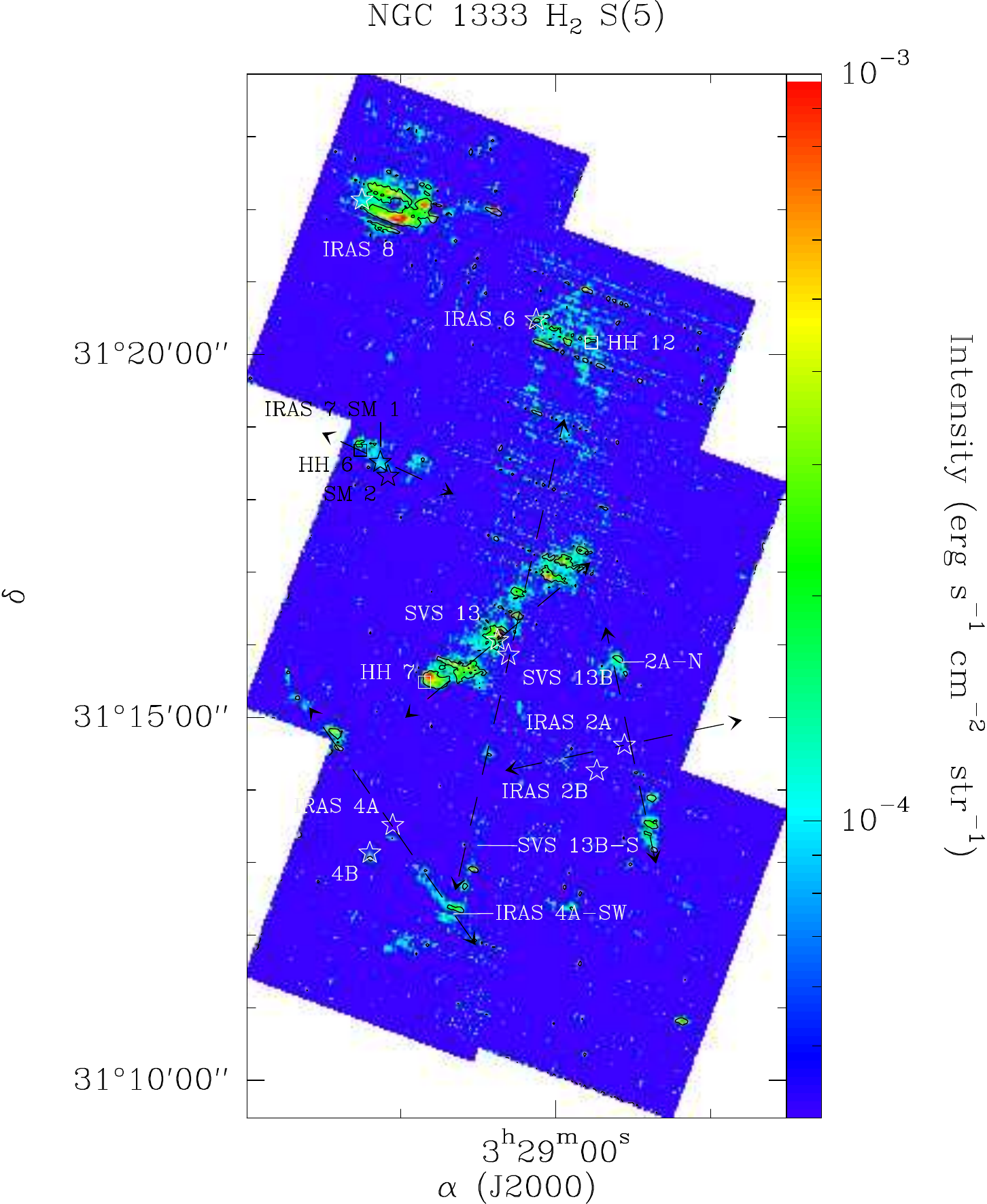}
  \caption{H$_{2}$ $S(5)$ map obtained with the SL module. Contours
    show the 3 $\sigma$ noise level. \label{fig:h2-s5-map}}
\end{figure}

\begin{figure}
  \figurenum{4~h}
  \centering \includegraphics[]{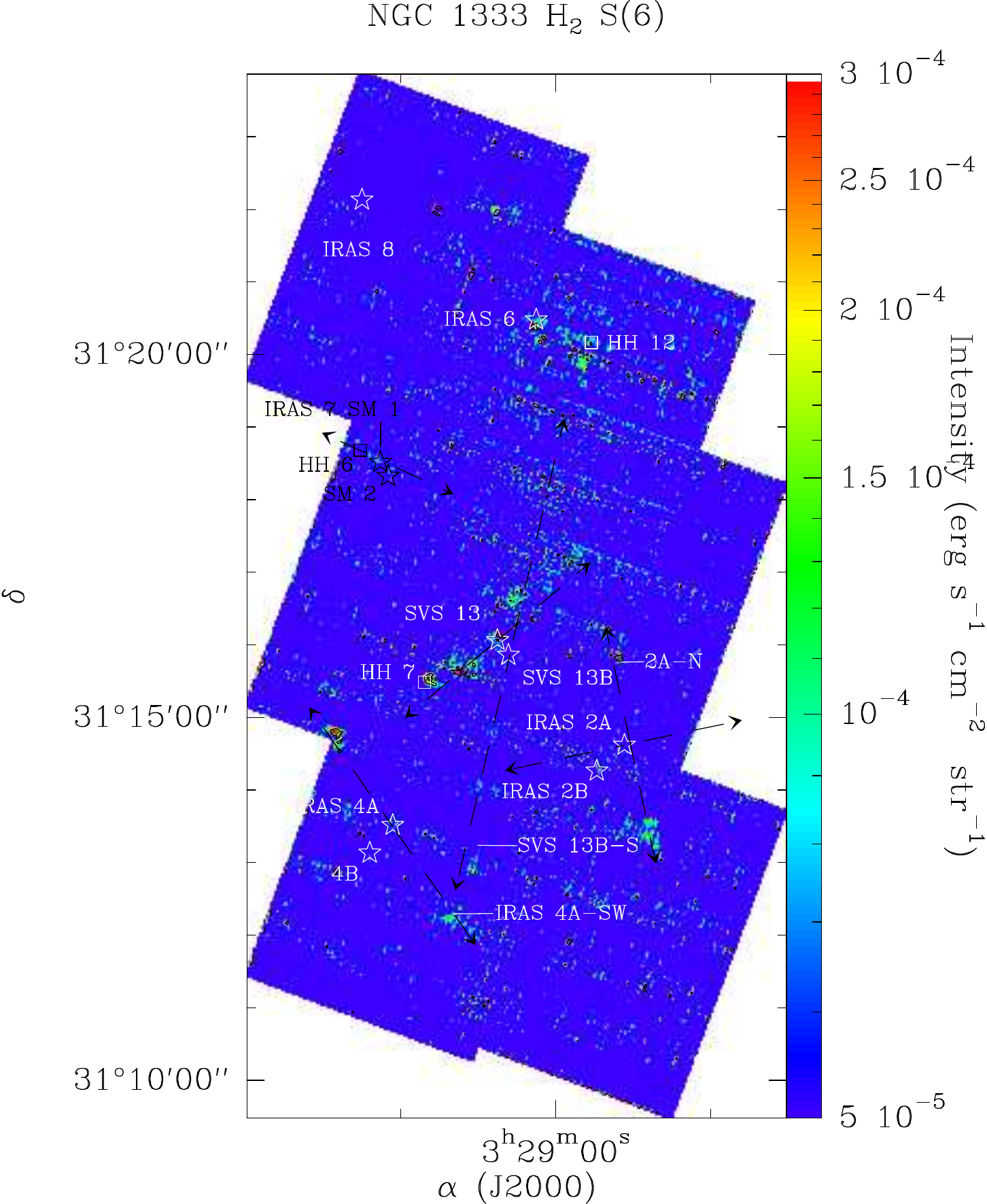}
  \caption{H$_{2}$ $S(6)$ map obtained with the SL module. Contours
    show the 3 $\sigma$ noise level. \label{fig:h2-s6-map}}
\end{figure}

\begin{figure}
  \figurenum{4~i}
  \centering \includegraphics[]{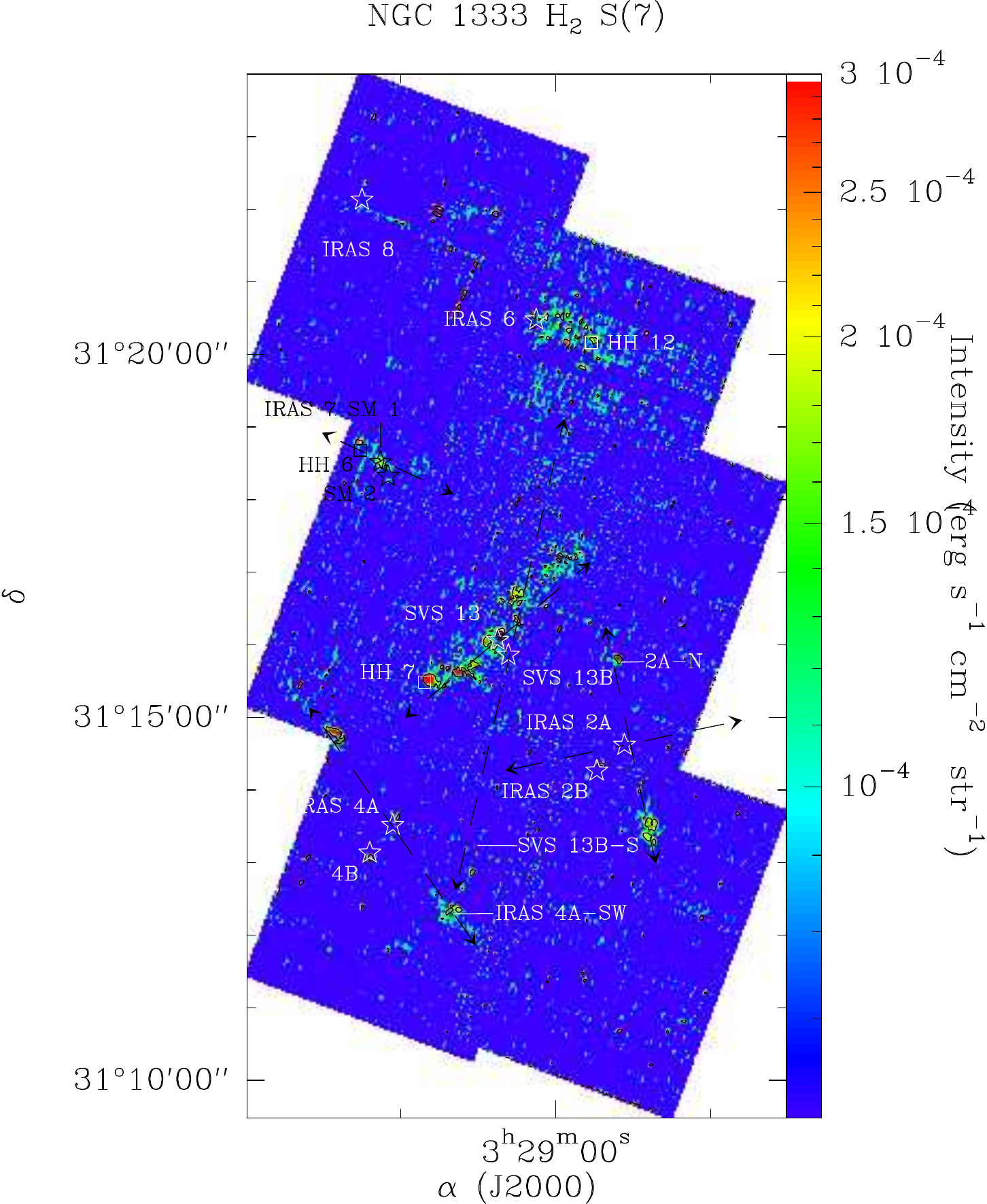}
  \caption{H$_{2}$ $S(7)$ map obtained with the SL module. Contours
    show the 3 $\sigma$ noise level. \label{fig:h2-s7-map}}
\end{figure}

\end{document}